\documentclass[final,3p,times,twocolumn]{elsarticle}

\usepackage{amssymb}
\usepackage{lineno}
\usepackage{subfig}
\usepackage{array}
\usepackage[export]{adjustbox}
\usepackage{float}
\usepackage{graphicx}
\usepackage{booktabs}
\usepackage{multirow}
\usepackage{amsmath}
\usepackage{xfrac}
\usepackage{commath}
\usepackage{bm}
\usepackage{color,soul}
\usepackage[labelsep=endash]{caption}
\usepackage{makecell}
\usepackage{enumitem}
\usepackage{placeins}
\usepackage{algorithm}
\usepackage{algpseudocode}
\usepackage{rotating}

\usepackage[table]{xcolor}
\colorlet{Gray1Fill}{black!10}
\colorlet{Gray2Fill}{black!20}
\colorlet{Gray3Fill}{black!40}

\begin{document}

\begin{frontmatter}



\title{Multi-physics inverse homogenization for the design of innovative cellular materials: application to thermo-mechanical problems}

\author[mecc]{Matteo Gavazzoni}
\author[mat]{Nicola Ferro}
\author[mat]{Simona Perotto}
\author[mecc]{Stefano Foletti}

\address[mecc]{Dipartimento di Meccanica, Politecnico di Milano \\ Via La Masa 1, Milano I-20156, Italy}

\address[mat]{MOX, Dipartimento di Matematica, Politecnico di Milano \\ Piazza L. da Vinci 32, Milano I-20133, Italy}

\begin{abstract}
We present a new algorithm to design lightweight cellular materials with required properties in a multi-physics context. In particular, we focus on a thermo-mechanical setting, by promoting the design of unit cells characterized both by an isotropic and an anisotropic behaviour with respect to mechanical and thermal requirements.  
The proposed procedure generalizes microSIMPATY algorithm to a multi-physics framework, by preserving all the good properties of the reference design methodology. The resulting layouts exhibit non-standard topologies and are characterized by very sharp contours, thus limiting the post-processing before manufacturing. The new cellular materials are compared with the state-of-art in engineering practice in terms of thermo-mechanical properties, thus highlighting the good performance of the new layouts 
which, in some cases, outperform the consolidated choices. 
\end{abstract}



\begin{keyword}

Topology optimization \sep Cellular materials \sep Multi-physics \sep Homogenization \sep Anisotropic mesh adaptation

\end{keyword}

\end{frontmatter}

\section{Introduction}\label{sec1}

Cellular materials represent an effective solution for structural applications where conventional monolithic materials fail to satisfy the design constraints \cite{Gibson1997}.
The fast advancements in additive manufacturing technologies, experienced in the last years, have further amplified the interest towards metamaterials. In addition, the possibility to employ a large variety of bulk materials in manufacturing processes (e.g., metals, polymers, ceramics~\cite{Rashed2016b,Bauer2014,Schaedler2016}) has enabled the design of new metamaterials, featuring innovative combinations of physical effective properties.
The possibility to blend different materials in order to reach diverse objectives proved to have a great impact
in all the contexts where multi-functionality is required.
For example, in~\cite{Ahmadi2014a,Yan2015,Taniguchi2016,Arabnejad2016a}, biocompatible 3D-printed metal bone implants promoting bone ingrowth are proposed by properly tailoring the material microstructure in order to reproduce the elastic modulus and the permeability of the human bone.
Other applications range from thermal-cloaking systems fitly combining microstructure geometry and orientation~\cite{Bandaru2015,Liu2018} to lattice-based heat exchangers, where good thermal conductivity and convection properties are exploited to enhance the devices' performance~\cite{Attarzadeh2021,Kaur2021}.

From a modeling viewpoint, the proposal of innovative multi-functional cellular materials can benefit from the most recent advancements in topology optimization~\cite{Bendsoe2004}, properly combined with direct and inverse homogenization processes~\cite{sigmund1994,andreassen2014determine,allaire19}.
Several optimization approaches can be exploited in the context of metamaterial design. The layout of the employed microstructures can be selected a priori, starting from consolidated dictionaries of unit cells~\cite{allaire19,vigliotti13,Wang2017b,Cheng2018,Panesar2018,Moussa2020}, or designed from scratch to match the expected effective properties~\cite{coelho08,naks13,blal2021,ferrer2016,Ferro2020,wccm21,auricchio20}.
In this context, a single- or a multi-objective topology optimization at the microscale can drive the design of new unit cells matching target properties at the macroscale, potentially in a multi-physics framework. For instance, 
the optimization of homogenized elastic properties is tackled in~\cite{Huang2011,Xia2014,Wang2014} with the aim of maximizing the bulk (or shear) modulus. To this aim, the authors control specific components of the homogenized elastic tensor or resort to the minimization of the compliance of a given structural part. Other works focus on a multi-physics optimization (for instance, by considering elastic, thermal and electrical properties) by providing microstructures optimized with respect to diverse objectives and physics~\cite{Torquato2003a,DeKruijf2007,Challis2008,Vineyard2021}.

Nevertheless, it is well-known that standard topology optimization techniques suffer from typical issues that may compromise the effective performance and manufacturability of the new layouts. Among the most recurrent, we mention the possible presence of intermediate densities, the non-smooth contours of the final design and the generation of unit cells which turn out to be unprintable since presenting too thin struts. All these drawbacks are strictly related to the selected computational grid: a coarse mesh promotes jagged boundaries and a diffused void/material interface; vice versa, an extremely fine mesh leads to an non affordable computational effort and fosters the generation of too complex structures.
Filtering offers a possible remedy to address all these concerns, by alternating smoothing with sharpening phases to be properly tuned. Such a tuning is not a trivial task and may often lead to non-optimal design solutions~\cite{DeKruijf2007,Bendsoe2004,SP98}.
\\
The selection of a computational mesh customized to the design problem has been proved to be instrumental in order to limit the main issues of topology optimization. For instance, in~\cite{soli2019}, the combination of a standard density-based method for topology optimization with an anisotropic mesh adaptation procedure has been used to get rid of intermediate densities, irregular boundaries and thin struts in the design of structures at the macroscale.
The proposed algorithm, named SIMPATY (SIMP with mesh AdaptiviTY), is based on a robust mathematical tool, namely an a posteriori estimator for the discretization error, and leads to final designs characterized by reliable mechanical properties as well as by free-form features.
The same procedure has been successfully exploited at the microscale, with the proposal of the microSIMPATY algorithm~\cite{Ferro2020}. So far, this procedure has been used for the design of unit cells with optimized mechanical properties in a linear elasticity setting~\cite{wccm21,Ferro2021}.

In this work, we propose a new pipeline for the design of new cellular materials, by extending the microSIMPATY algorithm to a multi-physics context. The objective is to obtain lightweight metamaterials with prescribed requirements on the elastic and thermal conductivity properties, characterized by a ready-to-print topology. 
The design strategy here developed 
is confined to a 2D setting and
has to be meant as a proof-of-concept, preliminary to a 3D implementation. However, to corroborate the effectiveness of the proposed methodology, we perform a cross-comparison between the new cells and the standard ones in thermo-mechanical applications.

The paper is organized as follows.
Section~\ref{method} represents the core of the paper. It provides the physical problem constraining the optimization process, the main theoretical tools to perform the optimization and formalizes the multi-physics design procedure in the MultiP-microSIMPATY algorithm. 
Three design cases are considered in Section~\ref{results} to challenge MultiP-microSIMPATY algorithm onto diverse multi-physics scenarios. Section~\ref{discussion} further analyzes the results in the previous section by 
comparing the new designs with the state-of-the-art. Finally, Section~\ref{conclusions} outlines the most remarkable contributions of the work together with some future perspectives.

\section{Methods}
\label{method}

In this paper, we refer to a multi-physics setting, where the standard linear elasticity equation
\begin{equation}\label{lin_eq}
-\nabla \cdot \sigma(\textbf{u}) = \textbf{f} \qquad \textrm{ in }\  \Omega \subset \mathbb{R}^2
\end{equation}
is combined with the thermal conduction problem
\begin{equation}\label{thermal_eq}
-\nabla \cdot \textbf{q}(\theta) = \textbf{h} \qquad \textrm{ in }\  \Omega \subset \mathbb{R}^2.
\end{equation}

The elasticity model is characterized by the stress tensor
\begin{equation}\label{voigt_elastic}
\begin{array}{ll}
\sigma(\textbf{u})=\begin{bmatrix} \sigma_{11}(\textbf{u}) \\ \sigma_{22}(\textbf{u}) \\ \sigma_{12}(\textbf{u}) \end{bmatrix} \hspace*{-.2cm} &= \begin{bmatrix} E_{1111} \ \ E_{1122} \ \ E_{1112} \\ E_{2211} \ \ E_{2222} \ \ E_{2212} \\ E_{1211} \ \ E_{1222} \ \ E_{1212} \end{bmatrix} \begin{bmatrix} \varepsilon_{11}(\textbf{u}) \\ \varepsilon_{22}(\textbf{u}) \\  2\varepsilon_{12}(\textbf{u}) \end{bmatrix}  \\[7mm] 
&= \textbf{E} \ \varepsilon(\textbf{u}),
\end{array}
\end{equation}
and by the force $\textbf{f}$ exerted on the body, 
where $\textbf{u} = [u_1, u_2]^{\textrm{T}}$ is the displacement field, $\varepsilon(\textbf{u}) = (\nabla \textbf{u} + \nabla \textbf{u}^T )/2$ is the small displacement strain tensor, and $\textbf{E}$ is the stiffness tensor characterizing the considered solid material. When dealing with homogeneous isotropic materials, tensor $\textbf{E}$ depends on the Lam\'e coefficients, $\lambda$ and $\mu$, functions of the Young modulus $E$ and the Poisson ratio $\nu$~\cite{gould}.

The thermal model \eqref{thermal_eq} is identified by the heat flux,
\begin{equation}
\begin{gathered}
\textbf{q}(\theta)=\begin{bmatrix} q_1(\theta) \\ q_2(\theta)  \end{bmatrix} = \begin{bmatrix} k_{11} \ \ k_{12} \\ k_{21} \ \ k_{22}  \end{bmatrix} \begin{bmatrix} \dfrac{\partial \theta}{\partial x_1} \\[12pt] \dfrac{\partial \theta}{\partial x_2} \end{bmatrix} = \textbf{k} \ \nabla \theta,
\end{gathered}
\label{voigt_thermal}
\end{equation}
with $\theta$ the temperature scalar field and $\textbf{k}$ the conductivity tensor of the solid material, and by the energy generation term
$\textbf{h}$. In particular, 
the diagonal entries in $\textbf{k}$ represent the material conductivities, while the off-diagonal terms are null.
\\
Equations \eqref{lin_eq} and \eqref{thermal_eq} are completed by suitable conditions which model the physical configuration along the boundary $\partial \Omega$ of the design domain $\Omega$.
\\
In standard structural optimization at the macroscale,
equations \eqref{lin_eq}-\eqref{thermal_eq} work as  constraints, after being properly modified to include a so-called design variable (we refer to function $\rho$ in the next section), possibly combined with additional design requirements.

Vice versa, when the optimization is applied to the microscale, it is crucial to properly transfer the physical characterization of the micro- to the macroscale, in order to make this two-scale computation feasible. In such a direction, direct and inverse homogenization represent widespread solutions~\cite{Hassani1998a,Hassani1998b,Terada2000,sigmund1994}.
The direct approach incorporates the microscopic effects into a homogenized macroscopic model, for instance, by means of an asymptotic expansion of the primal variable in terms of microscopic field fluctuations. As a consequence, the microscopic behaviour is known, whereas we have to identify the (homogenized) macroscopic characterization. In practice, this leads to modify the definition of the stress tensor and of the heat flux as 
$$\sigma^H(\textbf{u})= \textbf{E}^H \ \varepsilon(\textbf{u}), \quad \textbf{q}^H(\theta) = \textbf{k}^H \ \nabla \theta,$$
respectively where the homogenized stiffness tensor, $\textbf{E}^H$, and the homogenized thermal conductivity tensor, $\textbf{k}^H$, include the effects of the microscale. \\ 
On the contrary, inverse homogenization starts from desired macroscopic physical properties and designs the microscale in order to match such features, thus swapping the role played by known and unknown scales with respect to the direct homogenization, as detailed in the next section.

\subsection{Inverse homogenization}
Inverse homogenization is the procedure which allows us to design microstructures with prescribed properties at the macroscale. The required features are mathematically commuted into a goal functional $\mathcal{J}$ and into suitable constraints driving a topology optimization process~\cite{Ferro2020,Ferro2021,wccm21,huang11,noel2017}. In particular,
the optimization problem we are interested in reads
\begin{equation}
\displaystyle{\min_{\rho \in L^\infty \left(Y, [0, 1] \right)  } \mathcal{J}( z(\rho), \rho)}:
\begin{cases}
a_{\rho} \left( z(\rho), w \right) =F_{\rho}(w) \quad \forall w\in W \\[3pt]
{\bf L}_B \leq {\bf C}(z(\rho), \rho) \leq {\bf U}_B.
\end{cases}
\label{opt_problem_generic}
\end{equation}
The material distribution in the unit cell $Y$ at the micro-scale is modeled by means of the auxiliary scalar field $\rho$ that represents the relative material density, where it is assumed that $\rho=1$ labels the material, while $\rho=0$ identifies the void. However, since density $\rho\in  L^\infty \left(Y, [0, 1] \right)$ can take all the values in $[0, 1]$, it is standard to penalize the intermediate values (i.e., intermediate material densities) that are not physically consistent. To this aim, we resort to the SIMP method~\cite{Bendsoe2004}. In particular, in a linear elasticity setting, SIMP modifies the reference state equations by weighting the constitutive laws with a suitable power $\rho^p$ of the density.

The first constraint in \eqref{opt_problem_generic} models the physics of the problem. It coincides with the weak form of the state equation modified by the density function (subscript $\rho$ takes into account such a dependence), set in a suitable function space $W$~\cite{ern04}. The box inequality in \eqref{opt_problem_generic} drives the optimization process according to specified design and physical requirements, where vector ${\bf C}$ includes the quantities to be controlled through the corresponding lower and upper bounds, ${\bf L}_B$ a ${\bf U}_B$.

In the analysis below, we pick the objective functional $\mathcal{J}$ as
\begin{equation}
\mathcal{M}(\rho) = \int_Y \rho \ dY
\label{functional}
\end{equation}
since we are interested in minimizing the total mass, $\mathcal{M}$, of the cellular structure, i.e., to design lightweight materials.\\
In the context of the design of cellular materials with prescribed mechanical response, it is customary to choose as weighed state equation the elastic model at the microscale 
\begin{equation}\label{micro_elastic}
\begin{array}{lll}
&&a^{E,ij}_{\rho} \left( \textbf{u}^{*,ij}(\rho),\textbf{v} \right) = \dfrac{1}{\abs{Y}}\displaystyle\int_Y \rho^{p} \ \sigma(\textbf{u}^{*,ij}) : \varepsilon(\textbf{v}) dY  \\[3mm] 
&=&\dfrac{1}{\abs{Y}}\displaystyle\int_Y \rho^{p} \ \sigma(\textbf{u}^{0,ij}) : \varepsilon(\textbf{v}) dY =
F^{E,ij}_{\rho}(\textbf{v}),
\end{array}
\end{equation}
where $\textbf{u}^{*,ij}$ and ${\bf v}$ belong to the space $\mathcal{U}_{\#}^2= [H^1_{\circlearrowright}(Y)]^2$ of the $H^1(Y)$-functions satisfying periodic boundary conditions, and $ij \in I=\{11,22,12\}$.
According to a standard homogenization procedure, these equations model the $Y$-periodic displacement field fluctuations, $\textbf{u}^{*,ij}$, induced by the reference displacement fields, $\textbf{u}^{0,ij}$, with $\textbf{u}^{0,11}=[x, 0]^{\textrm{T}}$, $\textbf{u}^{0,22}=[0, y]^{\textrm{T}}$ and $\textbf{u}^{0,12}=[y, 0]^{\textrm{T}}$.\\
Since we are interested in a multi-physics inverse homogenization, we further constrain the topology optimization process with an additional weighed state equation. In particular, we consider the thermal conductivity model at the microscale 
\begin{equation}\label{micro_thermal}
\begin{array}{lll}
&&a^{k,m}_{\rho}  (\theta^{*,m}(\rho), v) = \displaystyle
\frac{1}{\abs{Y}}\int_Y \rho^s \ \textbf{q}(\theta^{*,m}) : \nabla v \ dY \\[3mm]
&=& \displaystyle \frac{1}{\abs{Y}}\int_Y \rho^s \ \textbf{q}(\theta^{0,m}) : \nabla v \ dY
 = F^{k,m}_{\rho}(v),
\end{array}
\end{equation}
where $\theta^{*,m}$ and $v\in \mathcal{U}_{\#}^1 = H^1_{\circlearrowright}(Y)$,  and $m\in J=\{1, 2\}$, and where index $s$ plays the same role as $p$ in \eqref{micro_elastic}. Analogously to \eqref{micro_elastic}, $\theta^{*,m}$ denotes the temperature fluctuations associated with the reference temperature fields $\theta^{0,m}$ (namely, $\theta^{0,1}=x$ and $\theta^{0,2}=y$).

The two problems at the microscale, \eqref{micro_elastic}
and \eqref{micro_thermal}, are instrumental to define the homogenized elastic tensor, ${\bf E}^H$, and the homogenized thermal conductivity tensor, ${\bf k}^H$, to be involved in the setting of the box constraints in \eqref{opt_problem_generic}. The component-wise definition of ${\bf E}^H$ and ${\bf k}^H$ is
\begin{equation}\label{elastic}
\begin{array}{ll}
E_{ijkl}^H=\displaystyle\frac{1}{\abs{Y}}&\displaystyle\int_Y \rho^p \left[ \sigma ( \textbf{u}^{0,ij} ) - \sigma (\textbf{u}^{*,ij}(\rho)) \right]  \\[5mm] & :\left[ \varepsilon(\textbf{u}^{0,kl}) - \varepsilon (\textbf{u}^{*,kl}(\rho)) \right] dY,
\end{array}
\end{equation}
\begin{equation}\label{thermal}
\begin{array}{ll}
k_{mn}^H=\displaystyle\frac{1}{\abs{Y}}&\displaystyle\int_Y \rho^s \left[ \textbf{q}(\theta^{0,m}) - \textbf{q} ( \theta^{*,m}(\rho) ) \right]  \\[5mm] & : \left[ \nabla \theta^{0,n} - \nabla  \theta^{*,n}(\rho)  \right] dY,
\end{array}
\end{equation}
respectively, with $ij, kl \in I$ and $m,n\in J$.\\
In particular, the two-sided inequality in \eqref{opt_problem_generic} will be exploited to promote diverse mechanical and thermal behaviours along the different spatial directions. To this aim, we constrain the two ratios $E_{2222}^H / E_{1111}^H$ and $k_{22}^H / k_{11}^H$ so that they vary in suitable ranges.
This choice allows us to penalize the mechanical and the thermal contributions in a different way along the two directions, as shown in the numerical assessment.
An additional two-sided control is enforced on the first and the last diagonal terms, $E_{1111}^H$ and $E_{1212}^H$, of the homogenized elastic tensor, as well as on the first diagonal term, $k_{11}^H$, of the homogenized thermal conductivity tensor.

To sum up, the optimization setting we are led to deal with coincides with the following problem:
\begin{equation}
\displaystyle{\min_{\rho \in L^\infty \left( Y, [0,1] \right)  } \mathcal{M}(\rho)}:
\begin{cases}
a^{E,ij}_{\rho} \left( \textbf{u}^{*,ij}(\rho),\textbf{v} \right) =F^{E,ij}_{\rho}(\textbf{v}) \\
\hspace*{2.8cm} \forall \textbf{v}\in\mathcal{U}^2_\#, ij \in I \\
a^{k,m}_{\rho} \left( \theta^{*,m}(\rho) ,v \right) = F^{k,m}_{\rho}(v) \\
\hspace*{2.8cm} \forall v\in\mathcal{U}^1_\#, m \in J \\
E_{1111}^{\textrm{low}} \leq E_{1111}^{H} \leq E_{1111}^{\textrm{up}} \\[5pt]
E_{1212}^{\textrm{low}} \leq E_{1212}^{H} \leq E_{1212}^{\textrm{up}} \\[5pt]
\left( \dfrac{E_{2222}}{E_{1111}} \right)^{\textrm{low}} \leq \dfrac{E_{2222}^H}{E_{1111}^H} \leq \left( \dfrac{E_{2222}}{E_{1111}} \right)^{\textrm{up}} \\[4mm]
k_{11}^{\textrm{low}} \leq k_{11}^{H} \leq k_{11}^{\textrm{up}} \\[5pt]
\left( \dfrac{k_{22}}{k_{11}} \right)^{\textrm{low}} \leq \dfrac{k_{22}^H}{k_{11}^H} \leq \left( \dfrac{k_{22}}{k_{11}} \right)^{\textrm{up}} \\[3mm]
\rho_{\textrm{min}} \leq \rho \leq 1
\end{cases}
\label{opt_problem}
\end{equation}
where all the bound values, $(\cdot)^{\rm low}$ and $(\cdot)^{\rm up}$, will be set according to the application at hand.
The last inequality in \eqref{opt_problem} is meant to ensure the well-posedness of both the elasticity and the thermal problems \eqref{micro_elastic} and \eqref{micro_thermal}, $\rho_{\min}$ being a suitable value in $(0,1)$ (see Section~\ref{results} for more details).

\subsection{Discretization on anisotropic adapted meshes}\label{mesh_adapt}

With a view to the solution of problem \eqref{opt_problem},
all the quantities involved in the state equations, as well as in the constraints, have to be discretized on a suitable tessellation of the unit cell $Y$.
For this purpose, we resort to a computational mesh $\mathcal{T}_h = \{K\}$ customized to the problem at hand and characterized by stretched elements (i.e., a so-called  anisotropic adapted mesh).
Mesh $\mathcal{T}_h$ is employed to discretize both the test and the trial functions in the state equations, as well as the density function $\rho$, by means of a finite element scheme~\cite{ern04}.
\\
The anisotropic reference setting is the one proposed in~\cite{formaggia2001}. 
In particular, the anisotropic features of each element $K$ coincide with the lengths, $\lambda_{1,K}$,  $\lambda_{2,K}$, and the directions, $\textbf{r}_{1,K}$, $\textbf{r}_{2,K}$, of the semi-axes of the ellipse circumscribed to $K$, through the standard affine map, $T_K:\hat K \to K$, between the reference element $\hat K$ and the triangle $K$.
\\
Concerning the adaptation procedure, we resort to a metric-based approach driven by an a posteriori estimator for the discretization error associated with the density function $\rho$. 
Among the error estimators available in the literature~\cite{AO,BangerthRannacher}, we refer to an a posteriori recovery-based error analysis. Following the seminal work by O.C. Zienkewicz and J.Z. Zhu~\cite{ZZ1987}, we control the $H^1$-seminorm of the discretization error on the density, $e_{\rho}=\rho-\rho_h$. The selection of such an estimator is motivated by the fact that the density $\rho$ exhibits strong gradients (i.e., large values for the $H^1$-seminorm) across the material-void interface. This feature will yield meshes whose elements are crowded along the boundaries of the structure, thus promoting the design of very smooth layouts. To this aim, we exactly integrate the so-called recovered error, $\textbf{E}_{\nabla} = P\left( \nabla \rho_h \right) - \nabla \rho_h$, namely,
\begin{equation}\label{error_approx}
\begin{array}{rl}
|e_{\rho}|^2_{H^1(Y)} &= \|\nabla e_{\rho}\|^2_{L^2(Y)} = \displaystyle \int_Y \abs{\nabla \rho - \nabla \rho_h}^2 dY \\[3mm]  
&\simeq \|\textbf{E}_{\nabla}\|^2_{L^2(Y)} = \displaystyle \int_Y |P\left( \nabla \rho_h \right) - \nabla \rho_h|^2 dY,
\end{array}
\end{equation}
where $\rho_h$ denotes the finite element discretization of $\rho$ in the space $V_h^r$ of the piecewise polynomials of degree $r\in \mathbb{N}$ associated with $\mathcal{T}_h$. The operator $P:[V_h^{r-1}]^2 \to [V_h^{s}]^2$ in \eqref{error_approx}, with $s \in \mathbb{N}$ , denotes the recovered gradient, which, in general, provides a more accurate estimate of the exact gradient $\nabla \rho$ with respect to the discrete gradient $\nabla \rho_h$. Several recipes are available in the literature to define $P$~\cite{zz3,rodriguez,maisano,wiberg}. In particular, we select operator $P: [V_h^0]^2 \to [V_h^0]^2$ as the area-weighted average of $\nabla \rho_h$ over the patch of the elements, $\Delta_K = \{ T \in \mathcal{T}_h: T \cap K \neq \emptyset \}$, associated with $K$, i.e., we opt for
\begin{equation}
P\left( \nabla \rho_h \right) (\textbf{x}) = \dfrac{1}{|\Delta_K|}\displaystyle\sum_{T \in \Delta_K} \abs{T} \ \nabla \rho_h \big \rvert_T \quad \forall \textbf{x} \in K,
\label{grad_reconstruction}
\end{equation}
with $|\omega|$ the area of the generic
domain $\omega \subset \mathbb{R}^2$, where we have set the degree of the finite element space for $\rho_h$ to $r = 1$. Space $V_h^1$ is also adopted to discretize the components of the displacement vectors $\textbf{u}^{*,ij}$ as well as the temperature fields $\theta^{*,m}$ in \eqref{opt_problem}, with $ij \in I$ and $m \in J$.

According to~\cite{enumath09,Farrell2011,farrell2011anisotropic}, we here adopt the anisotropic generalization of \eqref{error_approx}. This estimator essentially exploits the anisotropic counterpart of the definition of the $H^1$-seminorm~\cite{formaggia2001}, based on the symmetric semidefinite positive matrix $G_{\Delta_K}$, with entries 
\begin{equation}
\left[ G_{\Delta_K} (\nabla g) \right]_{i,j} = \displaystyle\sum_{T \in \Delta_K} \int_T \dfrac{\partial g}{\partial x_i}\dfrac{\partial g}{\partial x_j}\ dT \quad i,j = 1,2, 
\end{equation}
with $g \in H^1(Y)$, and where it is understood  $x_1 = x$ and $x_2 = y$.
Thus, the squared $H^1$-seminorm $|e_{\rho}|^2_{H^1(Y)}$ is evaluated by the (global) error estimator $\eta^2 = \sum_{K \in \mathcal{T}_h} \eta_K^2$, where 
\begin{equation}
\eta_K^2 = \frac{1}{\lambda_{1,K} \lambda_{2,K}} \displaystyle\sum_{i=1}^2 \lambda_{i,K}^2 \left( \textbf{r}_{i,K}^{\textrm{T}} \ G_{\Delta_K}(\textbf{E}_{\nabla}) \ \textbf{r}_{i,K} \right),
\label{disc_error}
\end{equation}
defines the local error estimator.
The contribution between brackets coincides with the projection of the squared $L^2$-norm of the recovered error along the anisotropic directions, while the scaling factor $(\lambda_{1,K}\lambda_{2,K})^{-1}$ guarantees the consistency with the isotropic case (for more details, see~\cite{enumath09}).

The new adapted mesh is generated after commuting the error estimator $\eta_K$ into a new mesh spacing (the metric), $\mathcal{M}$, consisting of the triplet $\{\lambda_{1,K}^{adapt}, \lambda_{2,K}^{adapt},  \textbf{r}_{1,K}^{adapt}\}$, where the direction $\textbf{r}_{2,K}^{adapt}$ is automatically defined being $\textbf{r}_{1,K}^{adapt} \cdot \textbf{r}_{2,K}^{adapt} = 0$, for each element $K \in \mathcal{T}_h$.
This operation is performed by taking into account three different criteria, namely, (i) the minimization of  the mesh cardinality $\# \mathcal{T}_h$; (ii) an accuracy requirement on the discretization error $|e_\rho|_{H^1(Y)}$ (i.e, on the error estimator $\eta$), controlled up to a user-defined tolerance \texttt{TOL}; (iii) the equidistribution of the error throughout the mesh elements (i.e., $\eta_K^2 = \texttt{TOL}^2/\# \mathcal{T}_h$).
These three criteria lead us to solve a constrained minimization problem on each triangle $K \in \mathcal{T}_h$. The solution to this local optimization problem can be analytically derived, as proved in~\cite{Bolla06}, being 
\begin{equation}
\begin{gathered}
\lambda_{1,K}^{adapt} = g_2^{-1/2} \left( \frac{\texttt{TOL}^2}{2 \ \#\mathcal{T}_h \ |\hat{\Delta}_K|} \right)^{1/2}, \quad \textbf{r}_{1,K}^{adapt} = \textbf{g}_2, \\
\lambda_{2,K}^{adapt} = g_1^{-1/2} \left( \frac{\texttt{TOL}^2}{2 \ \#\mathcal{T}_h \ |\hat{\Delta}_K|} \right)^{1/2}, \quad \textbf{r}_{2,K}^{adapt} = \textbf{g}_1
\end{gathered}
\label{adapt}
\end{equation}
where $g_1$, $g_2$ and $\textbf{g}_1$, $\textbf{g}_2$ are the eigenvalues and the eigenvectors of the scaled matrix $\hat{G}_{\Delta_{K}}(\textbf{E}_{\nabla}) = G_{\Delta_{K}}(\textbf{E}_{\nabla}) / |\Delta_{K}|$, with $g_1 \ge g_2 > 0$.
\\
Finally, the metric $\mathcal{M} = \{\lambda_{1,K}^{adapt}, \lambda_{2,K}^{adapt},  \textbf{r}_{1,K}^{adapt}\}_{K \in \mathcal{T}_h}$ has to be changed into a quantity associated with the vertices of $\mathcal{T}_h$, received as an input by the selected mesh generator.
A standard choice consists in an arithmetic mean formula applied to the patch of elements associated with each vertex in $\mathcal{T}_h$~\cite{Farrell2011,farrell2011anisotropic}.

The anisotropic mesh adaptation based on the metric \eqref{adapt} is customized to a topology optimization problem in the algorithm SIMPATY, proposed in \cite{soli2019}. This procedure has been successfully employed for the design of structures at the macroscale~\cite{soli2019,Ferro2020a,Ferro2020b}, as well as for the design of new metamaterials with the proposal of algorithm microSIMPATY~\cite{ferro2020density,wccm21}.
Moreover, a combination of topology optimization at the macro- and at the micro-scale is carried out  in~\cite{Ferro2021}. In particular, a multiscale topology optimization process is used for the design of orthotic devices for 3D printing manufacturing, with the proposal of patient-specific innovative solutions.
\\
It has been verified that the adoption of an adapted anisotropic mesh leads to free-form layouts characterized by very smooth boundaries both at the macro- and at the micro-scale, mitigating some of the well-known drawbacks of standard topology optimization, such as the massive employment of filtering, the staircase effect and the generation of too complex structures~\cite{DeKruijf2007,Bendsoe2004,SP98}. However, in~\cite{Ferro2020b} it has been observed that
the presence of deformed elements inside the structures makes the finite element analysis less reliable. To overcome this issue, the authors suggest a hybrid approach. Thus, the mesh is kept isotropic, with a uniform diameter $h^{\rm iso}$ in the full-material regions, $\{{\bf x} \in Y : \rho_h({\bf x}) > \rho^{\rm th}\}$ with $\rho^{\rm th}$ a user-defined threshold, whereas the stretched triangles are preserved along the material-void interface. Actually, these hybrid meshes ensure an effective balance between smoothness of the structure and robust engineering performances. For this reason, we resort to hybrid meshes in the sequel.

\subsection{Multi-physics optimization algorithm}
\label{algo_sect}
In this section we propose the multi-physics adaptive inverse homogenization procedure, which generalizes  the algorithm proposed in~\cite{ferro2020density}. 
The discretization of the state equations \eqref{lin_eq} and \eqref{thermal_eq} is performed with the open-source finite element solver FreeFEM \cite{FreeFem}, which provides the ideal environment to implement an anisotropic mesh adaptation procedure in Section~\ref{mesh_adapt} through the built-in mesh generator BAMG (Bidimensional Anisotropic Mesh Generator).

The developed multi-physics optimization procedure is 
listed in the pseudocode below.

\linespread{1.3}
\begin{algorithm}[h]
	\caption{MultiP-microSIMPATY}\label{algo}
	\begin{algorithmic}[1]
		\State {\bf Input}: \verb+CTOL+, \verb+kmax+, $\textbf{c}^{{l}}$, $\textbf{c}^{{u}}$, $\rho_h^0$, \verb+TOPT+, \verb+IT+, \verb+kfmax+, $\tau$, $\beta$, $\mathcal{T}^0_h$, \verb+TOL+, $\tt HYB$
		\State Set: \verb+k+ = 0, errC = 1+\verb+CTOL+;
		\While{errC $>$ \texttt{CTOL} \&  \texttt{k} $<$ \texttt{kmax}}
		\State $\rho_h^\texttt{k+1}$ = \verb+optimize+($\mathcal{J}$, $\mathcal{C}$, $\textbf{c}^{l}$, $\textbf{c}^{u}$, $\mathcal{G}$, $\rho_h^\texttt{k}$, \verb+TOPT+, \verb+IT+);
		\If{$\texttt{k} < \texttt{kfmax}$}
		\State $\rho_h^\texttt{k+1}$ = \verb+helmholtz+($\rho_h^\texttt{k+1}$, $\tau$);
		\State $\rho_h^\texttt{k+1}$ = \verb+heaviside+($\rho_h^\texttt{k+1}$, $\beta$);
		\EndIf
		\State $\mathcal{T}^{\texttt{k+1}}_h$ = \verb+adapt+($\mathcal{T}^{\texttt{k}}_h$, $\rho_h^{\texttt{k+1}}$,\verb+TOL+, $\tt HYB$);
		\State errC = $\abs{\#\mathcal{T}^{\texttt{k+1}}_h-\#\mathcal{T}^{\texttt{k}}_h}/\#\mathcal{T}^{\texttt{k}}_h$;
		\State \verb+k+ = \verb+k++1;
		\EndWhile
		\State $\mathcal{T}_h$ = $\mathcal{T}^{\texttt{k}}_h$;
		\State $\rho_h = \rho_h^\texttt{k}$;
		\State $\left[ \textbf{E}^H, \textbf{k}^H \right]$ = \verb+homogenize+($\rho_h$); \\
		\Return $\mathcal{T}_h$, $\rho_h$, $\textbf{E}^H$, $\textbf{k}^H$
	\end{algorithmic}
\end{algorithm}
\linespread{1}

The main loop (lines 3-12) includes an optimization step, a filtering phase and the mesh adaptation. 
At each global iteration \texttt{k}, the optimization problem is solved (line 4, function ${\tt optimize}$) by taking into account all the constraints on the components of the elastic and of the thermal conductivity tensors in \eqref{opt_problem}. To this aim, we use the interior point algorithm IPOPT~\cite{ipopt}, although any other optimization tool can be selected~\cite{mma}. IPOPT requires as input the functional $\mathcal{J}$ to be minimized; the vector $\mathcal{C}$ gathering the constrained quantities in the optimization procedure; the two vectors $\textbf{c}^{l}$ and $\textbf{c}^{u}$ of the lower and upper bounds for the components in $\mathcal{C}$; the array $\mathcal{G}$ collecting the derivative of the functional $\mathcal{J}$ and of the constraints $\mathcal{C}$ with respect to $\rho$, computed by the adjoint Lagrangian approach (for more details, we refer to~\cite{wccm21}); the initial guess $\rho_h^{\tt k}$ to start the optimization process;  the accuracy  $\tt TOPT$ for the minimization problem; the maximum number of iterations $\tt IT$ to stop the optimization.
In particular, in the numerical assessment of Section~\ref{results}, we set $\texttt{TOPT} = 10^{-5}$, and $\texttt{IT} = 100$ for $\texttt{k}=0$ and $\texttt{IT} = 10$ for all the successive iterations. The higher value for $\texttt{IT}$  for ${\tt k} = 0$ takes into account that the initial guess $\rho_h^{\tt 0}$ can be completely arbitrary with respect to the minimum to be reached.
On the contrary, a smaller value for ${\tt IT}$ is sufficient for ${\tt k} > 0$, since the initial guess, $\rho_h^{\tt k}$, coincides with the output of a previous optimization step.\\ 
Function $\tt optimize$ returns the density $\rho_h^{\texttt{k+1}}$ which is successively processed by means of a Helmholtz and a Heaviside filters (lines 6-7, functions $\tt helmholtz$ and $\tt heaviside$)~\cite{filtrati,filtratiB}. The two filtering operations work in a complementary way. The Helmholtz partial differential equation is instrumental to remove too thin features, although promoting intermediate densities along the layout contour. In more detail, it consists of a low-pass filter based on a diffusion kernel with radius $\tau \in \mathbb{R}^+$. On the contrary, the Heaviside filter, coinciding with a $\beta$-dependent regularization of the Heaviside function with $\beta \in \mathbb{R}^+$, penalizes the intermediate material densities, also due to the Helmholtz filter, thus increasing the sharpness of the material/void interface.
The combined filtering take place for the first $\tt kfmax$ global iterations only. This choice leads to start the mesh adaptation procedure with a density field which is free from too complex features, while exhibiting a clear alternation between void and material. The filtering phase becomes redundant when the optimization loop approaches the minimum, so that mesh adaptation alone suffices to ensure well-defined structures. In the next section, filtering parameters $\tau$ and $\beta$ are set equal to $0.02$ and $5$ respectively, while ${\tt kfmax} = 25$.\\
The next step coincides with the mesh adaptation procedure detailed in Section~\ref{mesh_adapt} and here represented by function $\tt adapt$ (line 9). The input parameter $\tt TOL$ establishes the accuracy of the error estimator $\eta$ through the predicted metric in \eqref{adapt}. Parameter $\tt HYB$ is a boolean flag that, in correspondence with the full material, switches the employment of an isotropic mesh on or off. 

The main loop is controlled by a check on the stagnation of the relative difference between the cardinality of two consecutive meshes (line 10), up to a maximum number of global iterations $\tt kmax$ (line 3).
The choices \texttt{TOL} $ = 10^{-5}$ and \texttt{kmax}$=100$ are preserved throughout all the numerical assessment below.

Algorithm MultiP-microSIMPATY returns the final adapted mesh $\mathcal{T}_h$, the optimized density $\rho_h$, the homogenized elastic and conductivity tensors, $\textbf{E}^H$ and $\textbf{k}^H$, computed by function $\tt homogenize$ (line 15), based on \eqref{elastic} and \eqref{thermal}.

We remark that the procedure itemized in Algorithm \ref{algo} is fully general and it
can be applied in a straightforward way to different multi-physics contexts after properly modifying the formulation in \eqref{opt_problem}.
\section{Results}
\label{results}
We analyze three different cases of microstructure design according to \eqref{opt_problem}. In order to highlight the interplay between the different (thermal and mechanical) physics involved, we consider configurations where the thermal conductivity and the elastic stiffness requirements act along different directions. For instance, a high shear stiffness combined with a high thermal conductivity along the $x$-direction orient the material along two opposite directions, with the prescription of a conflict configuration.

\begin{figure}[t]
	\hspace*{-4.mm}
	\includegraphics[width=0.5\textwidth]{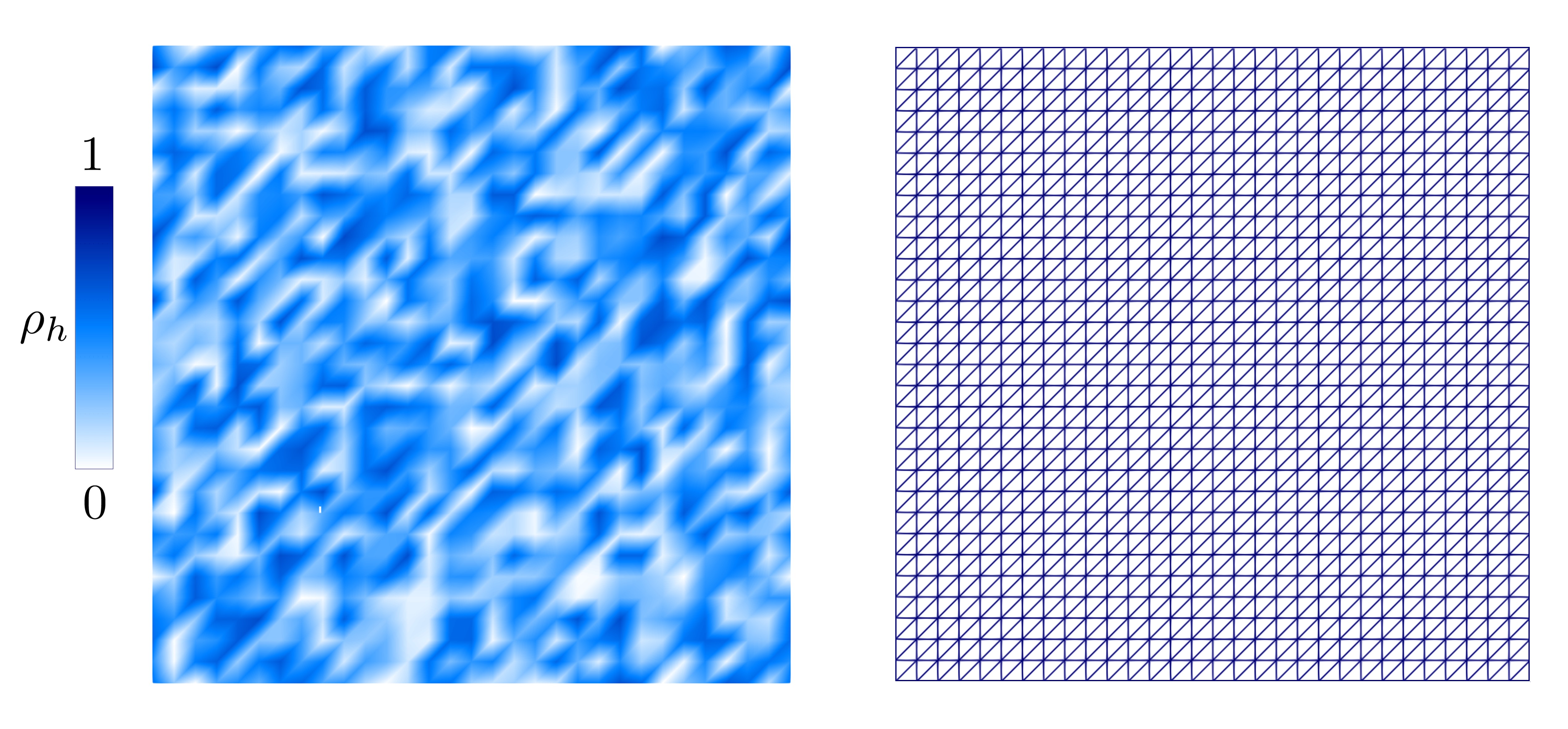}
	\caption{Initial guess $\rho_h^0$ (left) and corresponding mesh $\mathcal{T}_h^0$ (right).}
	\label{initial}
\end{figure}

The whole verification below shares common choices for some physical quantities and discretization parameters.
In particular, the unit cell $Y \subset \mathbb{R}^2$ is identified with the unitary square, $Y=(0,1)^2$.
Moreover, we set the Young modulus, $E$, and the Poisson ratio, $\nu$, to $1$ and $0.3$, respectively, and we consider an isotropic solid material with unitary thermal conductivity by setting $k_{11}=k_{22} = 1$. These choices allow us to obtain normalized homogenized mechanical and thermal properties for the cellular structures. Following~\cite{ferro2020density}, both the SIMP-powers, $p$ and $s$, in \eqref{elastic} and \eqref{thermal} are chosen equal to $4$ to penalize intermediate densities.
\\
Concerning the discretization frame, we choose a random density field, $\rho_h^{0}$ as the initial guess for the optimization process, defined on an initial structured mesh characterized by $30$ subdivisions per side, and with values ranging from $\rho_{min} = 10^{-4}$ to 1 (see Figure~\ref{initial} for an example).
Finally, to ensure a reliable finite element analysis, we resort to the hybrid mesh adaptation procedure ($\tt HYB$ = 1 in function $\tt adapt$).
In particular, we choose the threshold value $\rho^{\rm th} = 0.9$ to manage the alternation between isotropic and anisotropic elements, and the isotropic tessellation is characterized by the uniform diameter $h^{\rm iso} = 0.03$ (approximately $1/30$ of the design domain dimension).

After the optimization,
we perform a verification step to check the actual mechanical and thermal properties of the material yielded by a periodic repetition of the optimized unit cell.
To this aim, we use the Abaqus software\footnote{Abaqus, Dassault Syst{\`e}mes Simulia Corp, United States.}. The layouts provided by Algorithm~\ref{algo} are imported in Abaqus after a thresholding which neglects the density smaller than 0.75. The obtained geometry is remeshed on a uniform isotropic triangular mesh with an average size equal to $0.01$, while the displacement and temperature fields are discretized with quadratic finite elements, completed with periodic boundary conditions. The verification here performed can be considered as a preliminary step towards the integration of MultiP-microSIMPATY algorithm into a common workflow for structural analysis.

\subsection{Design case 1}
The main goal of this first optimization process is to design a lightweight unit cell characterized by isotropic mechanical homogenized properties and, vice versa,  anisotropic thermal homogenized features. 
This problem can be cast in setting \eqref{opt_problem}, after making the following choices for the constraints:
\begin{equation}\label{design1}
\hspace*{-3.5cm}
\left\{
\begin{array}{l}
0.05 \leq E_{1111}^{H} \leq 0.08  \\[5pt]
0.055 \leq E_{1212}^{H} \leq 0.080  \\[5pt]
1 \leq \dfrac{E_{2222}^H}{E_{1111}^H} \leq 2  \\[4mm]
0.01 \leq k_{11}^{H} \leq 1.00  \\[5pt]
0.00 \leq \dfrac{k_{22}^H}{k_{11}^H} \leq 0.58. 
\end{array}
\right.
\end{equation}
The isotropic mechanical behaviour and the anisotropic thermal properties are enforced by the constraints in \eqref{design1}$_3$ and \eqref{design1}$_5$.
In particular, we expect ratios $E_{2222}^H / E_{1111}^H$ and $k_{22}^H / k_{11}^H$ to coincide with the corresponding lower and upper bounds, respectively.
Moreover, since a control on the ratios does not ensure $E_{1111}^H$, $E_{2222}^H$, $k_{11}^H$, and $k_{22}^H$ to be in a physically admissible range of values, we further constrain the optimization through the box inequalities \eqref{design1}$_1$ and \eqref{design1}$_4$. Finally, a control on the component $E_{1212}^H$ of the homogenized stiffness tensor closes the minimization problem, thus further restricting the solution space.

\begin{figure}[h]
	\centering
	    \includegraphics[height=5.75cm]{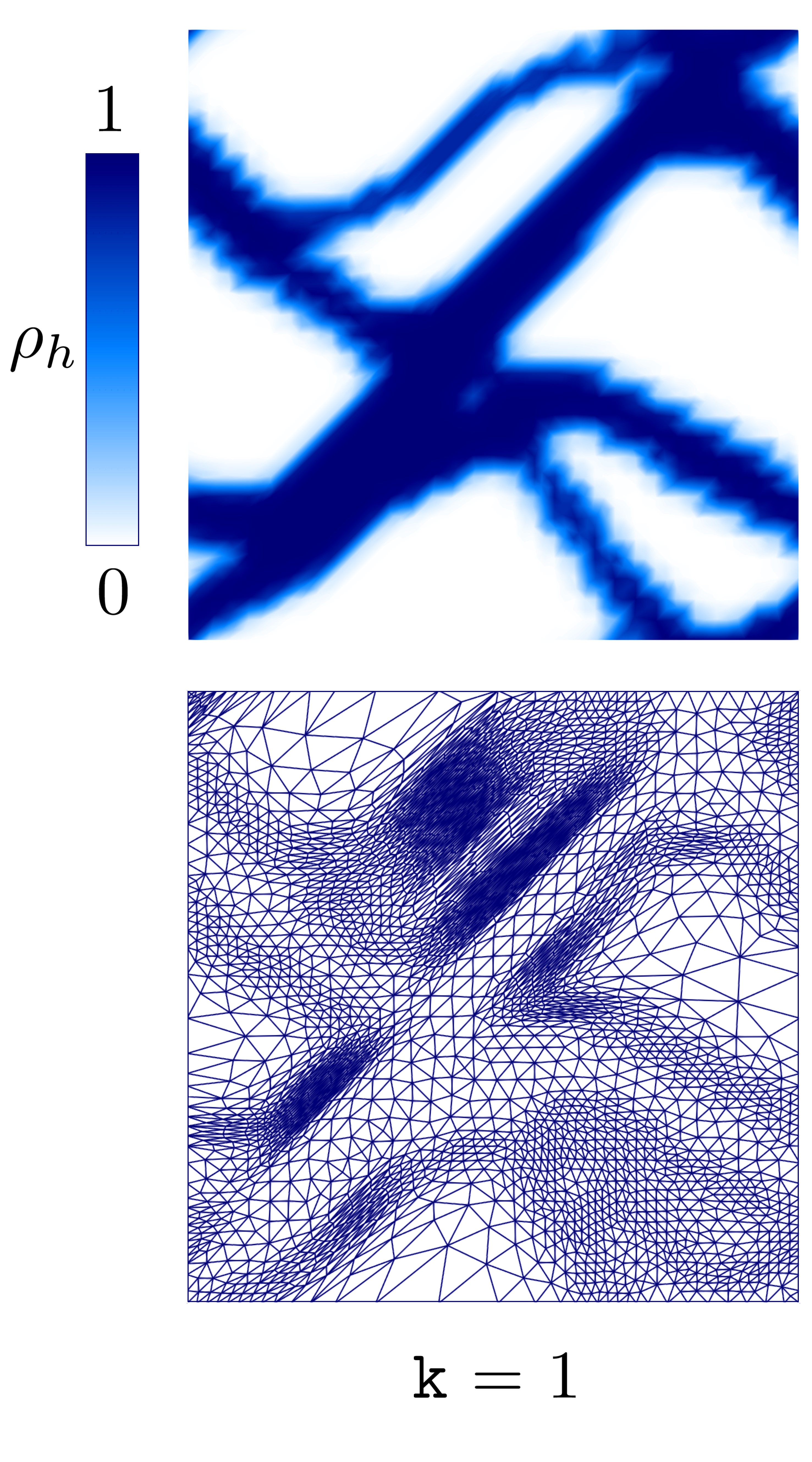}
		\hspace{-0.42cm}
		\includegraphics[height=5.75cm]{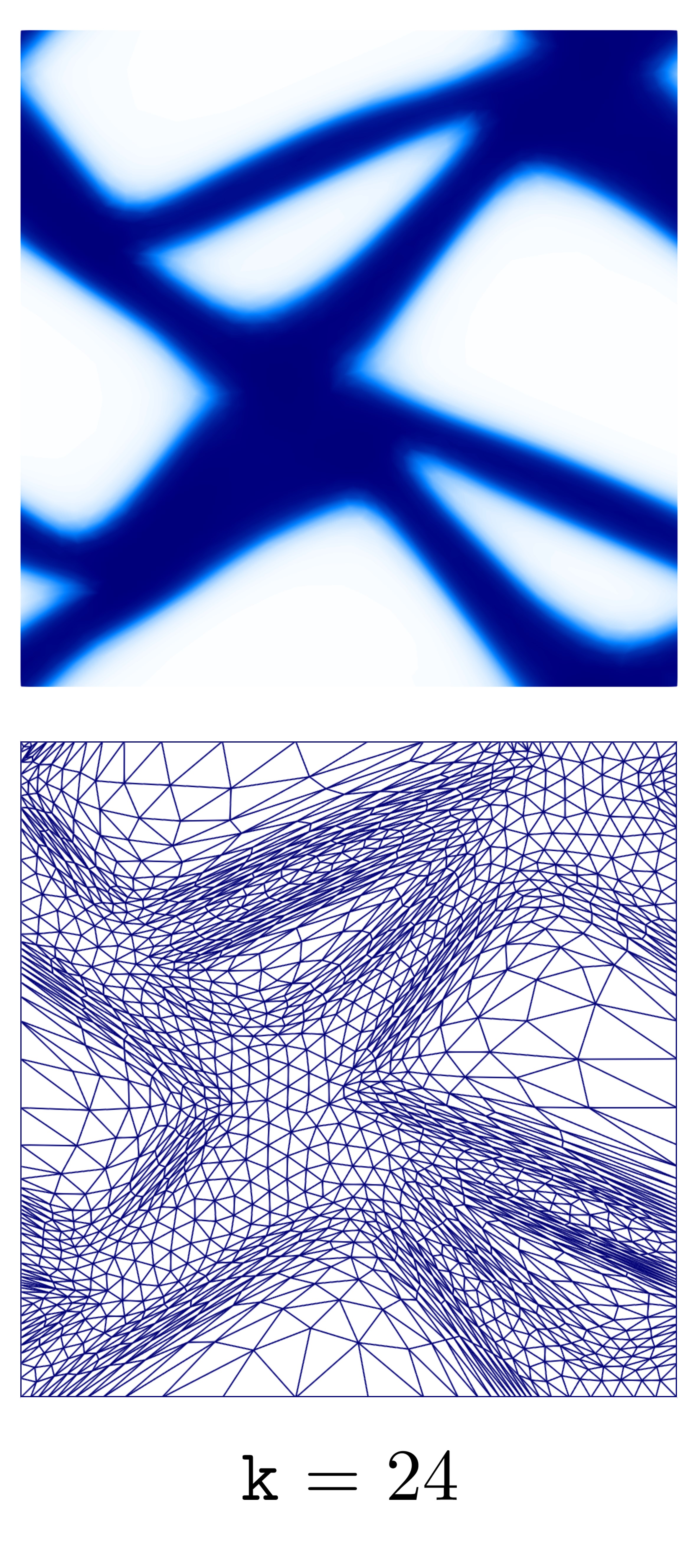}
		\hspace{-0.42cm}
		\includegraphics[height=5.75cm]{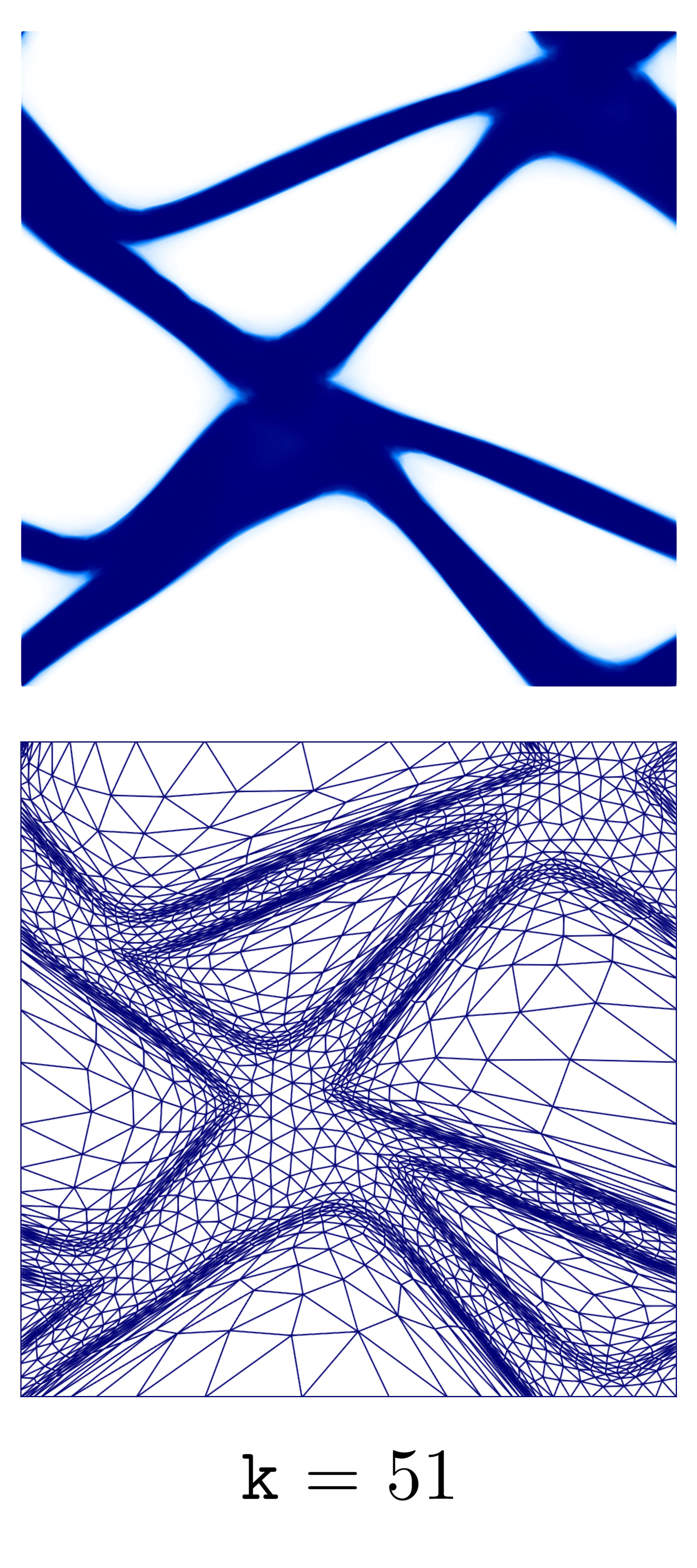}
	\caption{Design case 1: density field (top) and associated anisotropic adapted mesh (bottom) for three different global iterations.}
	\label{cell1_k}
\end{figure}

For the values set for the input parameters, MultiP-microSIMPATY algorithm converges in $51$ global iterations. Figure~\ref{cell1_k} shows the layout and the associated anisotropic adapted mesh at three different iterations.

\begin{table}[b!]
	\begin{center}
		\caption{Design cases 1, 2, 3: values of the constraints and of the objective functional computed with Abaqus software, together with the lower and the upper bounds, $c^l$ and $c^u$, involved in the optimization.}
		\label{eff_prop}
		\begin{tabular}{lccccc|c}
			\toprule
			& $E_{1111}^H$ & $E_{1212}^H$ & $\frac{E_{2222}^H}{E_{1111}^H}$ &  $k_{11}^H$ & $\frac{k_{22}^H}{k_{11}^H}$ & $\mathcal M$
			\\
			\midrule
			\rowcolor{Gray1Fill}
			\multicolumn{7}{c}{Design case 1} \\
			\hline
			\rowcolor{Gray1Fill}
			$c^{u}$ & 0.080 & 0.080 & 2.000 & 1.000 & 0.580 & \\[2pt]
			\rowcolor{Gray1Fill}
			$c$ & 0.038 & 0.056 & 1.299 & 0.199 & 0.566 & \\[2pt]
			\rowcolor{Gray1Fill}
			$c^{l}$ & 0.050 & 0.055 & 1.000 & 0.010 & 0.000 & \multirow{-3}{*}{0.292}\\
			\hline
			\rowcolor{Gray2Fill}
			\multicolumn{7}{c}{Design case 2} \\
			\hline
			\rowcolor{Gray2Fill}
			$c^{u}$ & 0.350 & 0.150 & 2.000 & 1.000 & 2.000 & \\[2pt]
			\rowcolor{Gray2Fill}
			$c$ & 0.250 & 0.086 & 0.299 & 0.317 & 0.597 & \\[2pt]
			\rowcolor{Gray2Fill}
			$c^{l}$ & 0.230 & 0.080 & 0.300 & 0.300 & 0.000 & \multirow{-3}{*}{0.412} \\
			\hline
			\rowcolor{Gray3Fill}
			\multicolumn{7}{c}{Design case 3} \\
			\hline
			\rowcolor{Gray3Fill}
			$c^{u}$ & 0.150 & 0.100 & 1.100 & 0.400 & 1.100 &  \\[2pt]
			\rowcolor{Gray3Fill}
			$c$ & 0.151 & 0.083 & 1.074 & 0.260 & 1.002 & \\[2pt]
			\rowcolor{Gray3Fill}
			$c^{l}$ & 0.100 & 0.080 & 1.000 & 0.250 & 1.000 & \multirow{-3}{*}{0.415}\\
			\bottomrule
		\end{tabular}
	\end{center}
\end{table}

\noindent
We remark that the final topology of the layout is already detected at the first iteration, although the quality of the solution is improved throughout the optimization process.
In particular, at the first iteration ($\texttt{k}=1$), we observe a significant staircase effect together with the presence of intermediate densities along the micro-structure interface. At the end of the filtering phase (${\tt k} = 24$), the jagged boundaries are fully smoothed, despite the intermediate densities still blur the design. The spreading effect along the material/void interface is gradually reduced when switching off the filtering, i.e., for ${\tt k} > 24$, as shown by the last column in Figure~\ref{cell1_k}. Thus, 
the final optimized solution ($\texttt{k}=51$) shows an extremely sharp transition from material to void and smooth boundaries, which make the structure ready for printing or manufacturing, with a limited need for post-processing.
\\
Concerning the adapted mesh, we recognize the effect of the hybrid approach, which combines stretched elements to discretize the strong gradients of the density field, coarse anisotropic triangles outside the structure, isotropic elements in correspondence with the material. 

\begin{figure}[h]
\centering
	\includegraphics[width=0.48\textwidth]{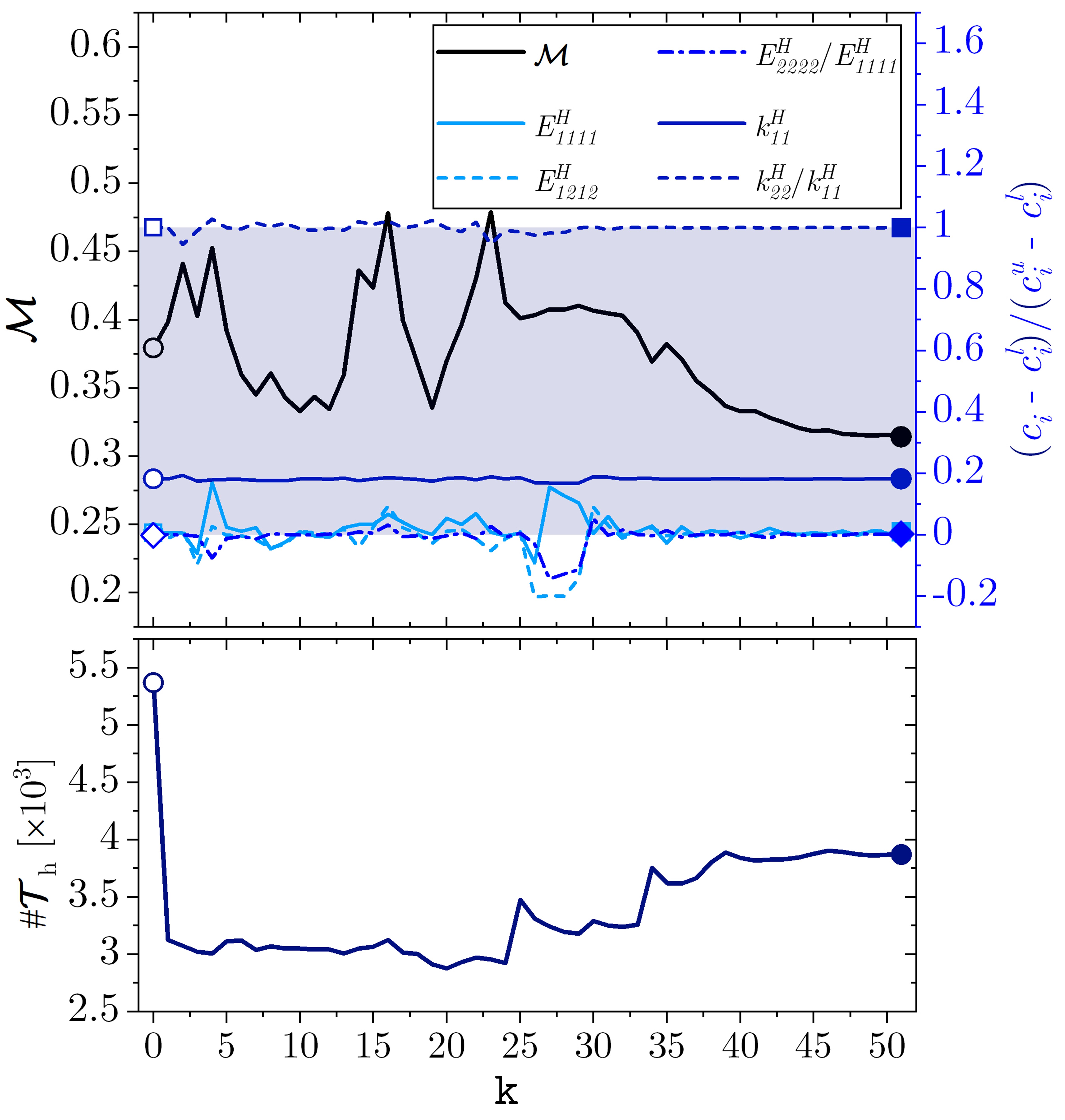}
	\caption{Design case 1: evolution of the objective functional $\mathcal{M}$ and of the constraints $c_i$ (top); trend of the mesh cardinality $\#\mathcal{T}_h$ (bottom) throughout the global iterations \texttt{k}.}
	\label{cell1_plot}
\end{figure}

Additional quantitative information on the MultiP-microSIMPATY algorithm is provided by Table~\ref{eff_prop} and by the diagrams in Figure~\ref{cell1_plot}, which show the evolution of the objective function and of the constrained quantities (top), together with the trend of the mesh cardinality (bottom), over the global iterations.

Notice that the values of the constraints have been normalized between 0 and 1 (see the highlighted area in the top panel of Figure~\ref{cell1_plot}). It is evident that the mass exhibits a completely different trend when compared with the constrained quantities.
The value of the objective function oscillates with values between 0.325 and 0.475 over the first $35$ iterations, and eventually converges towards a stable phase. On the contrary, all the constrained quantities are characterized by mild oscillations.
In particular, $k_{11}^H$ remains essentially constant over the whole optimization process. 
The plot of the ratios $E_{2222}^H / E_{1111}^H$ and $k_{22}^H / k_{11}^H$ confirms that the two inequalities are in conflict so that the active constraints are the lower and upper bound, respectively.
Moreover, from the values in Table~\ref{eff_prop}, it can be observed
that the stiffness component along the $x$-direction, $E_{1111}^H$, reaches a value which is about 25\% lower than the corresponding $c^l$. This can be ascribed to the presence of very thin struts generated by the severe thresholding ($\rho_h < 0.75$) applied before performing the analyses in Abaqus.

The evolution of the topology in Figure~\ref{cell1_k} is consistent with the trend in Figure~\ref{cell1_plot} (top panel). The topology does not essentially vary during the optimization process, according to the almost constant trend of the constraints. On the other hand, the highly oscillatory trend of $\mathcal M$ in the first optimization stage is related to the effect of the smoothing and of the sharpening operations which are confined to the first $24$ iterations. 
From ${\tt k} = 25$, only the minimization process and the mesh adaptation contribute to a mass variation, with less striking changes.

\begin{figure}[t!]
	\centering
	\includegraphics[height=2.35cm]{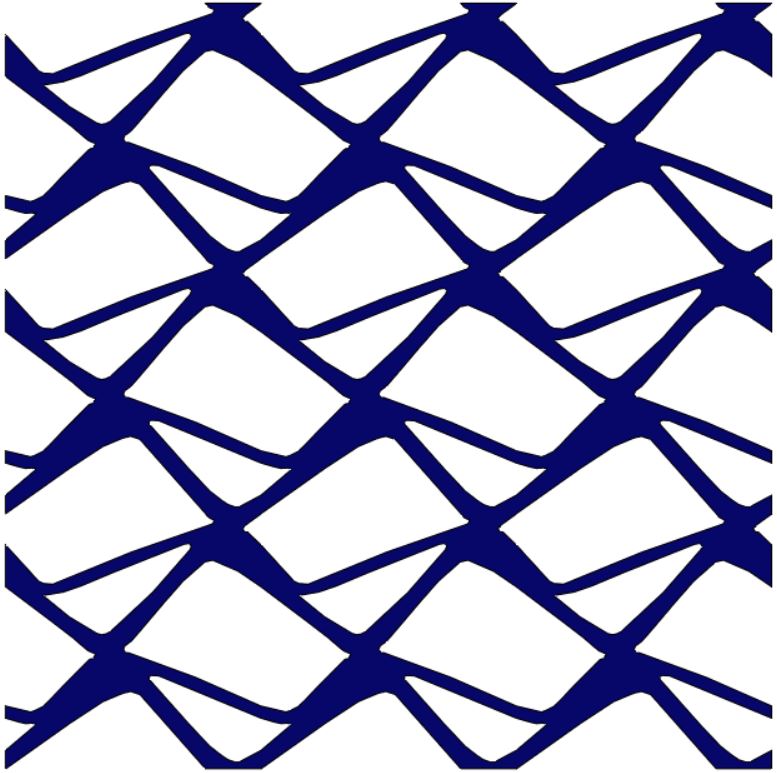}
	\hspace{0.15cm}
	\includegraphics[height=2.35cm]{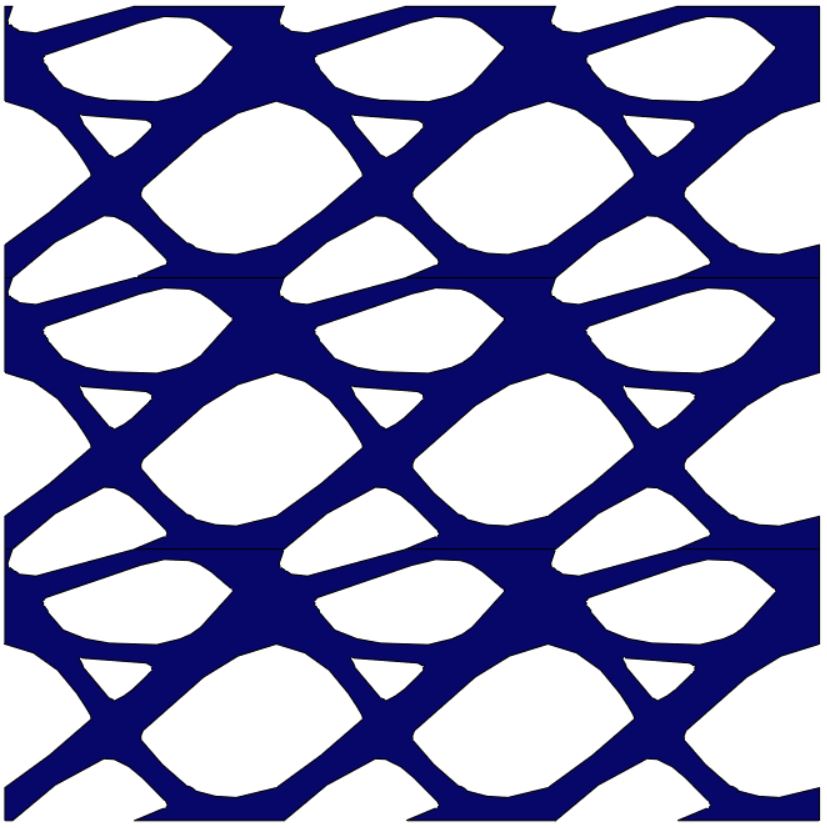}
	\hspace{0.15cm}
	\includegraphics[height=2.35cm]{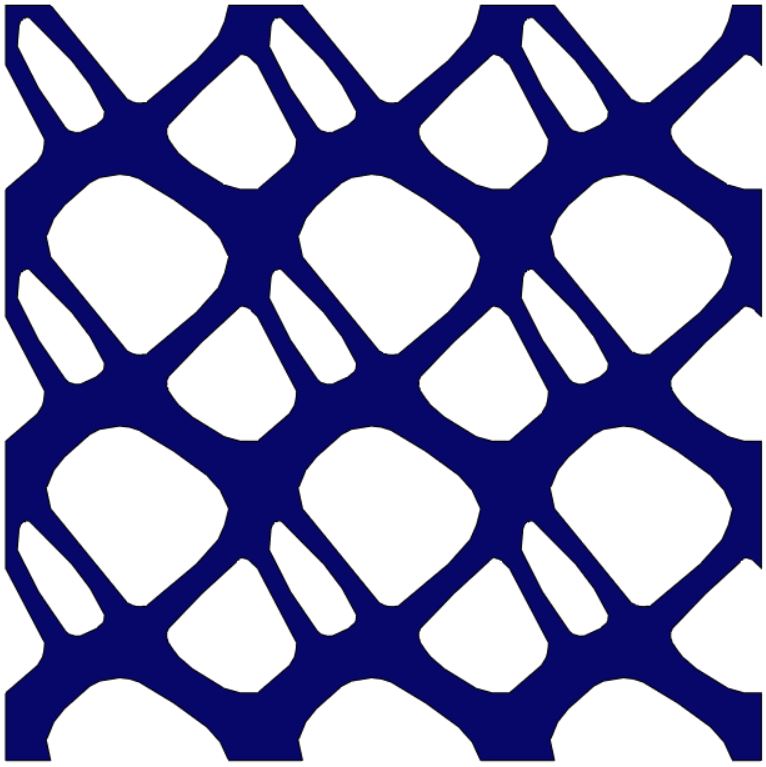}
	\caption{Design cases 1, 2, 3 (left-right): 
	$3 \times 3$-cell meta-material.}
	\label{RVEs}
\end{figure}

Finally, in Figure~\ref{RVEs} (left) we show the $3 \times 3$-cell material generated by a periodic repetition of the optimized unit cell.

\subsection{Design case 2}
The second MultiP-microSIMPATY run aims at designing a microstructure that provides high stiffness and thermal conductivity along the $x$-direction and a high shear stiffness. As for the Design case 1, these requirements might originate a set of conflicting constraints. In fact, the two former demands are expected to orient the material along the $x$-direction, while the latter requirement prescribes also the presence of material along the diagonal of the cell $Y$, which could react by tension to shear loading. 
This design setting is formalized by problem \eqref{opt_problem} when completed by the following set of constraints:

\begin{equation}\label{design2}
\hspace*{-3.5cm}
\left\{
\begin{array}{l}
0.23 \leq E_{1111}^{H} \leq 0.35  \\[5pt]
0.08 \leq E_{1212}^{H} \leq 0.15  \\[5pt]
0.3 \leq \dfrac{E_{2222}^H}{E_{1111}^H} \leq 2.0  \\[4mm]
0.3 \leq k_{11}^{H} \leq 1.0  \\[5pt]
0 \leq \dfrac{k_{22}^H}{k_{11}^H} \leq 2. 
\end{array}
\right.
\end{equation}
We highlight that the bounds for the stiffness tensor components to be promoted, $E_{1111}^H$ and $E_{1212}^H$, are set by taking into account the mass minimization goal, i.e., by keeping them considerably lower than 1. 
\begin{figure}[t]
	\centering
	\includegraphics[height=5.75cm]{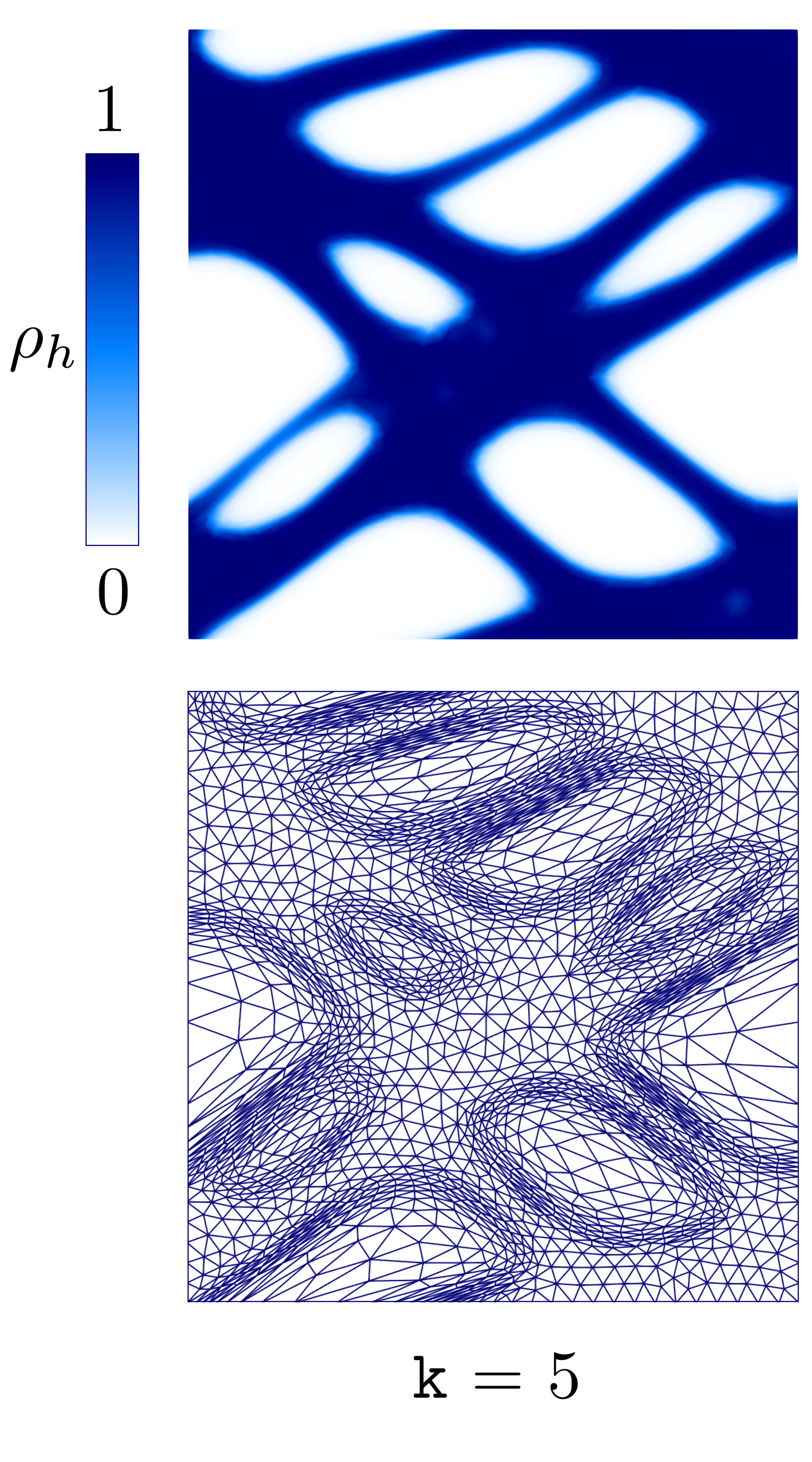}
	\hspace{-0.42cm}
	\includegraphics[height=5.75cm]{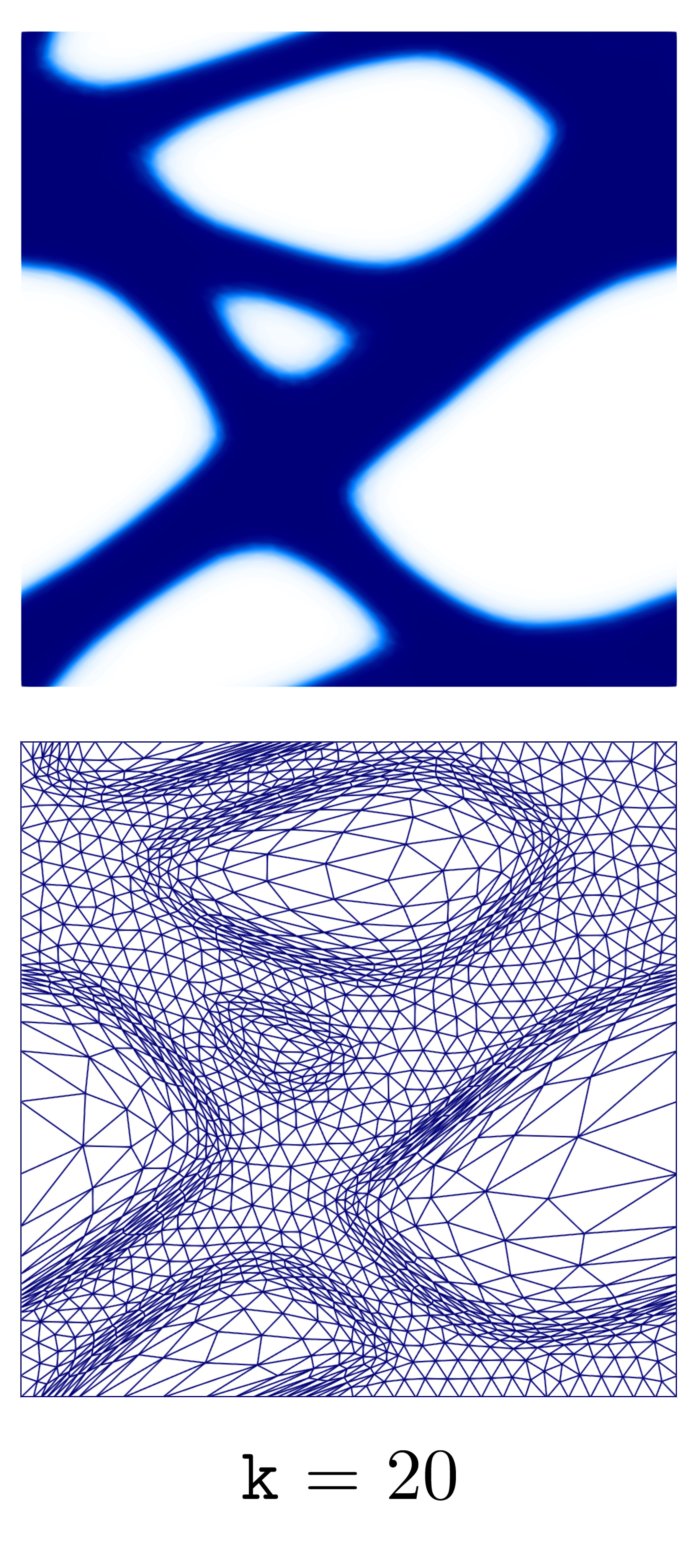}
	\hspace{-0.42cm}
	\includegraphics[height=5.75cm]{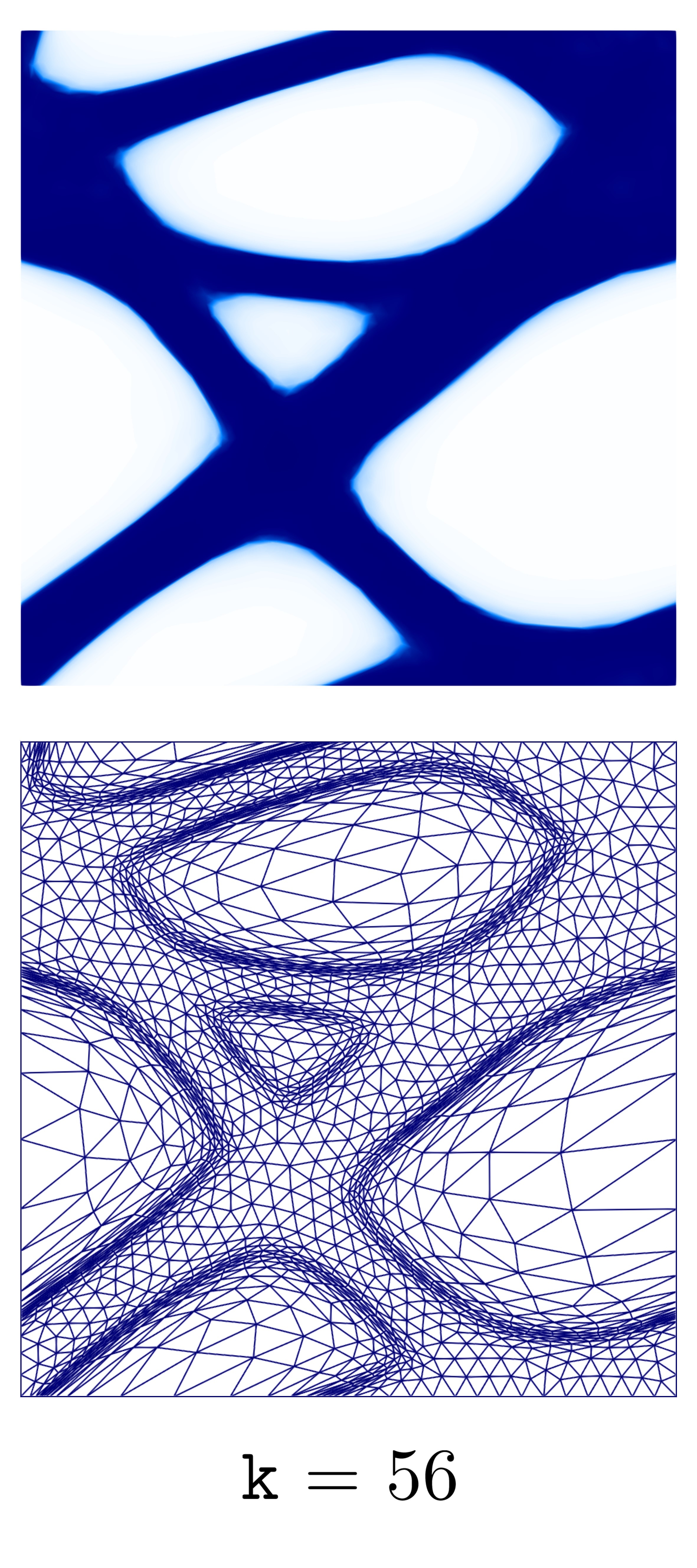}
	\caption{Design case 2: density field (top) and associated anisotropic adapted mesh (bottom) for three different global iterations.}
	\label{cell2}
\end{figure}

Algorithm~\ref{algo} stops in $56$ iterations due to mesh stagnation.
Figure~\ref{cell2} gathers the density field distribution together with the associated anisotropically adapted computational mesh at iterations ${\tt k} =$ 5, 20, 56.
At the fifth iteration the cell presents very thin struts that are progressively erased by the combined action of the Helmholtz and the Heaviside filters. For ${\tt k} = 20$, the topology essentially coincides with the final optimized one, although the layout still exhibits intermediate density values along the boundaries.
The structure contours become sharper and sharper throughout the next iterations when filtering is switched off and thanks to the mesh adaptation procedure. 

Concerning the final topology, we observe that most of the material is aligned along the two main diagonals of $Y$. This guarantees high shear stiffness, while ensuring a low stiffness along the $y$-direction, so that the lower bound for $E_{2222}^H / E_{1111}^H$ is reached. 
On the other side, the requirements on $E_{1111}^H$ and $k_{11}^H$ are taken into account by the two thinner struts along the $x$-direction, which improve the corresponding stiffness and the thermal conductivity.
Figure~\ref{RVEs} (center) provides a sketch of the metamaterial associated with the optimized cell in a $3 \times 3$ cellular pattern.

For a more quantitative characterization of the optimized structure in terms of mass and reached constraints, we refer to Table~\ref{eff_prop}. 
We notice that, to address the conflict among the several requirements, the optimization process pushes all the constrained quantities towards the lower bound of the corresponding range, while increasing the mass of the structure if compared, for instance, with the previous design case.

\subsection{Design case 3}
As a third design, we carry out the optimization of a microcell characterized by similar stiffness and thermal conductivity along the $x$- and $y$-directions and by a high shear stiffness. This leads to solve problem \eqref{opt_problem} when the following constraints are enforced:
\begin{equation}\label{design3}
\hspace*{-3.5cm}
\left\{
\begin{array}{l}
0.10 \leq E_{1111}^{H} \leq 0.15  \\[5pt]
0.08 \leq E_{1212}^{H} \leq 0.10  \\[5pt]
1.0 \leq \dfrac{E_{2222}^H}{E_{1111}^H} \leq 1.1  \\[4mm]
0.25 \leq k_{11}^{H} \leq 0.40  \\[5pt]
1.0 \leq \dfrac{k_{22}^H}{k_{11}^H} \leq 1.1. 
\end{array}
\right.
\end{equation}
The limited range for the two ratios $E_{2222}^H / E_{1111}^H$ and $k_{22}^H/k_{11}^H$ is consistent with the request for comparable stiffness and thermal conductivities along the two directions, whereas the mass minimization goal justifies the tight variation for the other tensors components.

\begin{figure}[h]
	\centering
	\includegraphics[height=5.75cm]{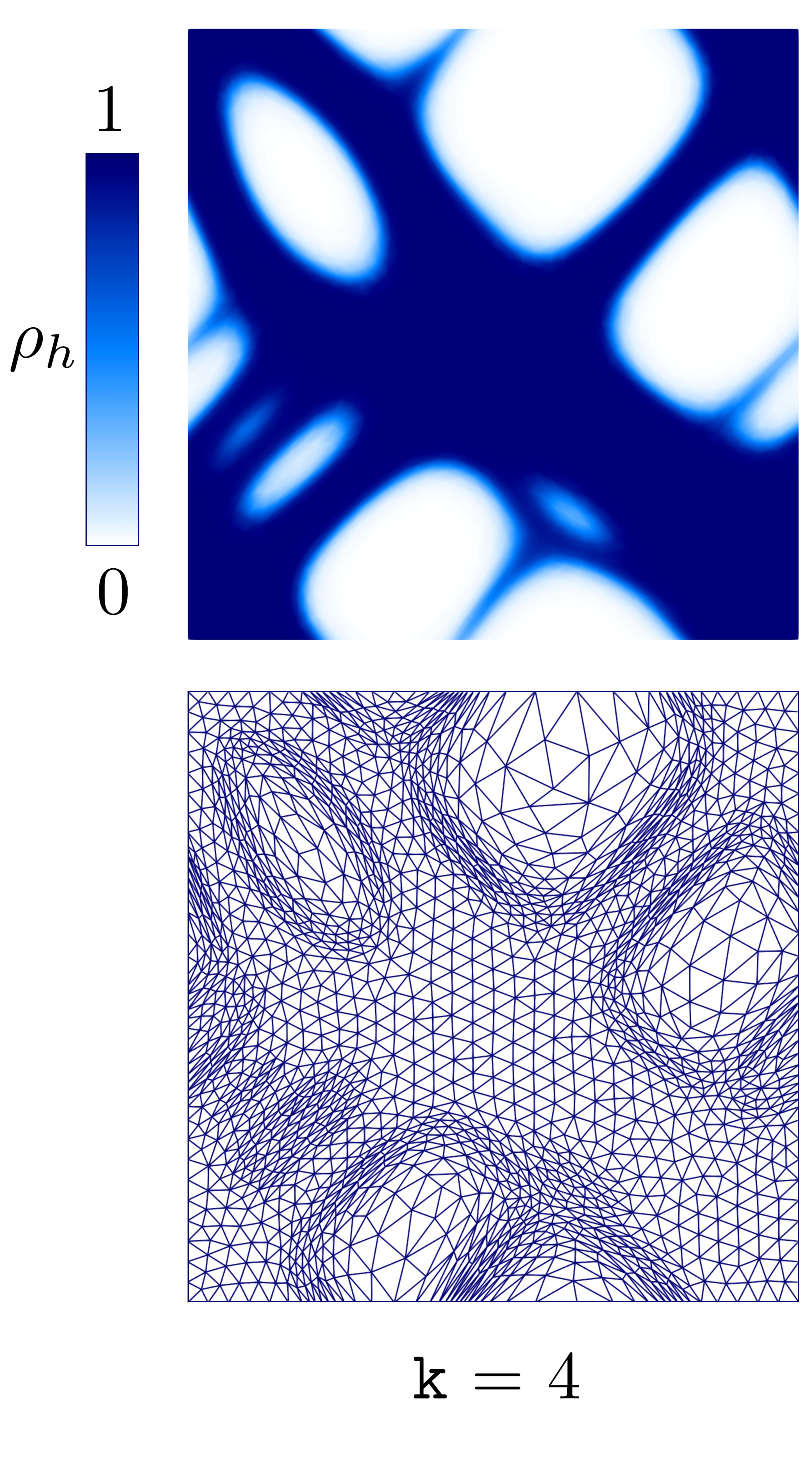}
	\hspace{-0.42cm}
	\includegraphics[height=5.75cm]{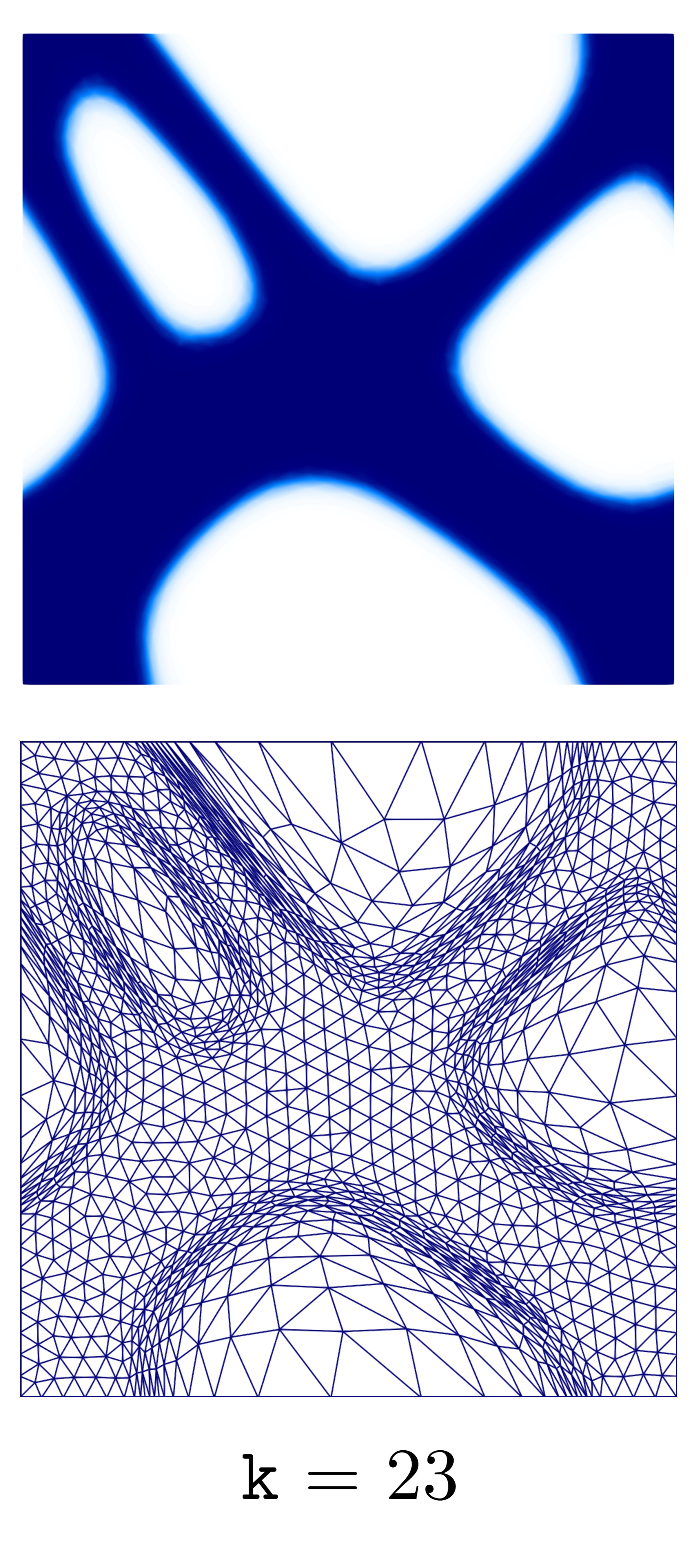}
	\hspace{-0.42cm}
	\includegraphics[height=5.75cm]{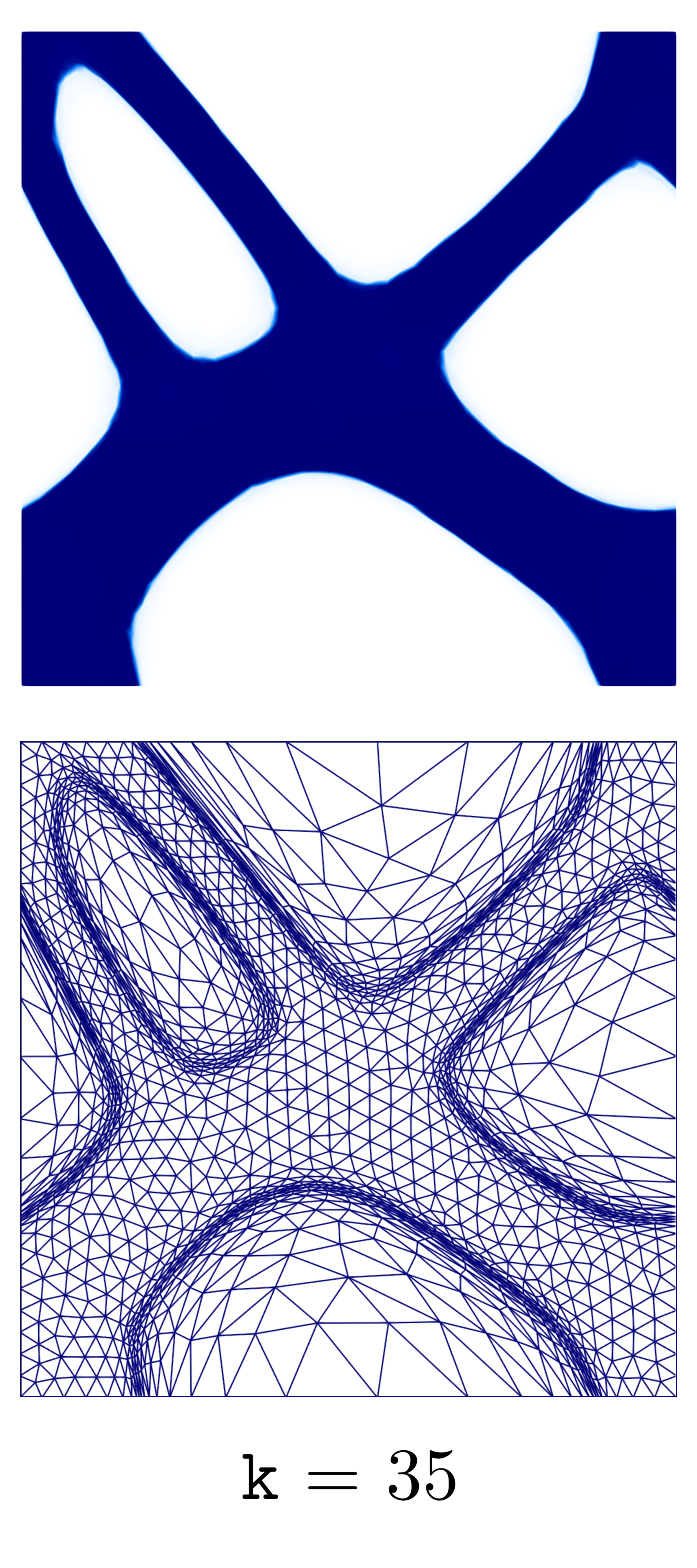}
	\caption{Design case 3: density field (top) and associated anisotropic adapted mesh (bottom) for three different global iterations.}
	\label{cell3}
\end{figure}

MultiP-microSIMPATY algorithm resorts to $35$ loops before satisfying the stopping criterion.
Figure \ref{cell3} shows the density field and the mesh for three different global iterations of the algorithm. As for the previous design cases, thin features are removed by filtering during the first $24$ iterations, while intermediate densities are erased in the second part of the process by the mesh adaptation procedure. As a consequence, the final micro-structure exhibits very sharp density gradients, so that little post-processing has to be applied.
In the final layout, most of the material is allocated along the two main diagonals of the domain, which ensures the required high shear stiffness as well as the balance between stiffness and thermal conductivity with respect to the horizontal and vertical directions.

Table~\ref{eff_prop} offers some additional quantitative information regarding the optimized structure. All the box constraints are satisfied (with a slight violation for the component $E_{1111}^H$), in the presence of a structure mass comparable with the one obtained for the Design case 2 (about $40\%$ with respect to the full material configuration).
We refer to Figure~\ref{RVEs} (right) for an example of the microcellular material associated with the optimized cell.

\section{Discussion of results}
\label{discussion}
This section is meant to highlight the benefits led by MultiP-microSIMPATY algorithm. To this aim, we compare the layouts provided by the proposed methodology with unit cells available in engineering practice and with cellular materials designed by a standard inverse homogenization procedure, which does not exploit mesh adaptation.

\subsection{Comparison with off-the-shelf designs}
This first investigation is carried out 
by comparing each of the three designs in the previous section with state-of-the-art unit cells 
in terms of mechanical and thermal performance, after setting a reference value for the overall mass. The quantities involved in such a comparison are
the homogenized elastic modulus, $E_x^H$ and $E_y^H$, associated with the direction $x$ and $y$, which
coincide with 
the inverse of the diagonal entries, $C_{11}^H$ and $C_{22}^H$, of the compliance matrix ${\textbf{C}^H} = ({\textbf{E}}^H)^{-1}$; 
the homogenized shear modulus,
$G^H,$ equal to the inverse of the third diagonal entry of matrix ${\textbf{C}}^H$;
the homogenized thermal conductivities, $k_{11}^H$ and $k_{22}^H$, along the $x$- and $y$-direction.
The results of this analysis are summarized in Table~\ref{Tab2_comparison}. 
\begin{table}[t!]
	\begin{center}
		\caption{Comparison between 
		the MultiP-microSIMPATY optimized structures and off-the-shelf designs in terms of homogenized elastic and thermal properties, for comparable volume fraction values.}
		\label{Tab2_comparison}
		\newcolumntype{C}{ >{\centering\arraybackslash} m{1.25cm} }
		\newcolumntype{D}{ >{\centering\arraybackslash} m{0.7cm} }
		\newcolumntype{N}{ >{\raggedleft} m{0.2cm} }
		\begin{tabular}{N C D D D D D}
			\toprule
			& & $E_{x}^H$ & $E_{y}^H$ & $G^H$ & $k_{11}^H$ & $k_{22}^H$  \\
			\midrule
			\rowcolor{Gray1Fill}
			\multicolumn{7}{c}{Design case 1} \\
			\hline
			\rowcolor{Gray1Fill}
			\vspace{2pt}
			D1 & \includegraphics[width=1.2cm]{immagini/cella1_RVE.jpg} & 0.012 & 0.015 & 0.056 & 0.200 & 0.113 \\[2pt]
			\hline
			\rowcolor{Gray1Fill}
			A &
			\includegraphics[width=1.25cm]{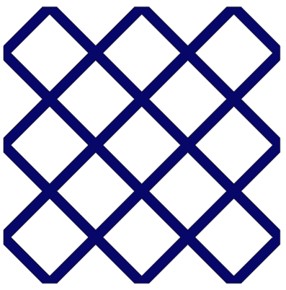} & 0.009 & 0.009 & 0.075 & 0.163 & 0.163 \\[2pt]
			\rowcolor{Gray1Fill}
			B &
			\includegraphics[width=1.25cm]{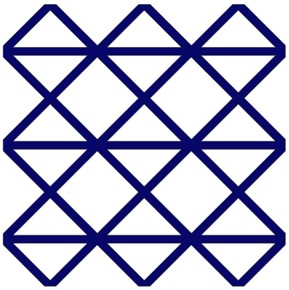} & 0.095 & 0.042 & 0.059 & 0.198 & 0.131 \\
			\hline
			\rowcolor{Gray2Fill}
			\multicolumn{7}{c}{Design case 2} \\
			\hline
			\rowcolor{Gray2Fill}
			\vspace{2pt}
			D2 &
			\includegraphics[width=1.2cm]{immagini/cella2_RVE.jpg} & 0.126 & 0.039 & 0.082 & 0.317 & 0.126 \\[2pt]
			\hline
			\rowcolor{Gray2Fill}
			C &
			\includegraphics[width=1.18cm]{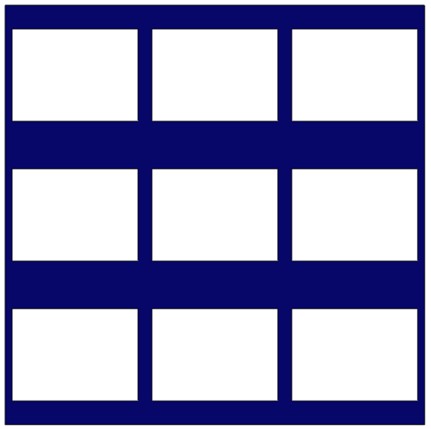} & 0.341 & 0.116 & 0.002 & 0.432 & 0.125 \\
			\hline
			\rowcolor{Gray3Fill}
			\multicolumn{7}{c}{Design case 3} \\
			\hline
			\rowcolor{Gray3Fill}
			\vspace{2pt}
			D3 &
			\includegraphics[width=1.2cm]{immagini/cella3_RVE.jpg} & 0.070 & 0.070 & 0.082 & 0.260 & 0.261 \\[2pt]
			\hline
			\rowcolor{Gray3Fill}
			L &
			\includegraphics[width=1.2cm]{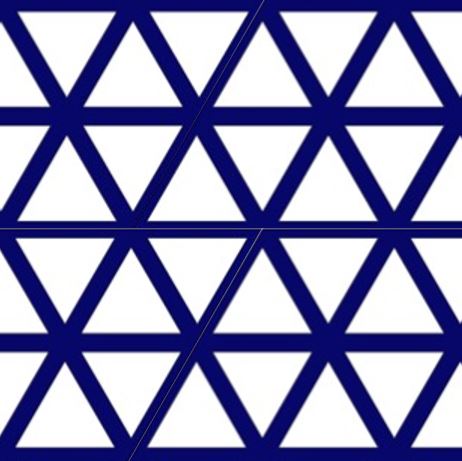} & 0.188 & 0.188 & 0.072 & 0.255 & 0.255 \\
			\bottomrule
		\end{tabular}
	\end{center}
\end{table}

Concerning the Design case 1, we perform two comparisons. Since the geometry provided by MultiP-microSIMPATY is similar to a square cell rotated by 45$^{\circ}$, we choose simple squares (A and B) characterized by the same rotation as state-of-the-art unit cells.
The basic squares in layout A fully couple mechanical and thermal features, thus excluding this cell for the purpose addressed in the first design case.
This justifies the selection of cell B where the reinforcing horizontal strut
mimics the very thin diagonal member connecting the adjacent sides in the proposed layout (D1). 
From a structural perspective, the horizontal strut in B increases the nodal connectivity and reacts with tension/compression to a load applied along the $x$-axis. This fact is confirmed by the non-isotropic elastic behaviour of the material (compare the values $E_x^H$ and $E_y^H$). Regarding thermal conduction, the strut promotes heat transfer along the horizontal direction, as highlighted by the discrepancy between ${k_{11}^H}$ and ${k_{22}^H}$. In the optimized layout D1, the thin member is instead slightly inclined and does not connect two opposite nodes. Thus, the elastic modulus along the two directions is similar since the strut reacts by bending to a load applied along the $x$-axis. Moreover, the thin member promotes the heat transfer along the $x$-direction, thus decoupling the ratios ${E_{y}^H}/{E_{x}^H}$ and ${k_{22}^H}/{k_{11}^H}$.

The unit cell D2 has been designed to ensure high stiffness and conductivity along the $x$-direction, as well as a high shear modulus.
As reference layout, we consider a square cell characterized by a rectangular cavity. This choice offers us a trivial solution to optimize stiffness and conductivity along direction $x$.
The optimization performed by MultiP-microSIMPATY is corroborated by the values of $G^H$. In fact, cell D2 is characterized by a shear modulus which is approximately $40$ times higher when compared with the reference layout, although 
the values of $E_x^H$ and $k_{11}^H$ for cell D2 are, on average, 30\% lower with respect to cell C.

Finally, the design case D3 aims at ensuring equal elastic modulus and conductivity along the $x$- and $y$-directions, as well as a high shear modulus.
The paradigm for an isotropic stretch-based lattice, namely the standard triangular cell (L), is assumed as the off-the-shelf layout.
A comparison between the corresponding values in Table~\ref{Tab2_comparison} shows a $15$\% increment in the shear modulus of cell D3. In addition, both cells D3 and L exhibit the requested isotropic behaviour in terms of the selected mechanical and thermal properties.

\subsection{Comparison with standard inverse homogenization}
This section is meant to verify the benefits led by mesh adaptation in the context of multi-physics inverse homogenization, in accordance with the preliminary remarks in Section~\ref{results}. \\
To this aim, we carry out a comparison between MultiP-microSIMPATY algorithm and a standard inverse homogenization procedure. This comparison is performed in terms of mass. 
We expect that the employment of mesh adaptation 
leads to efficiently allocate the available material, thus promoting the mass minimization.
As a reference standard approach, we implement a non-adaptive version of Algorithm~\ref{algo}, where the adaptation loop (lines 3-12) is replaced by the single call 
$$
\rho_h = \verb+optimize+(\Tilde{\mathcal{J}}, \Tilde{\mathcal{C}}, \textbf{c}^{l}, \textbf{c}^{u}, \Tilde{\mathcal{G}}, \rho_h^\texttt{0}, \verb+TOPT+, \verb+IT+);
$$
We refer to this variant of Algorithm~\ref{algo} as to MultiP-microSIMP.
In this case, the optimization is performed on the filtered density, so that the goal functional, the constraints and the associated derivatives are modified accordingly (this justifies the new notation $\mathcal{Q} \rightarrow \Tilde{\mathcal{Q}}$, with $\mathcal{Q} = \mathcal{J}, \mathcal{C}, \mathcal{G}$, where $\Tilde{\mathcal{Q}}$ refers to quantities dependent on the filtered density).
This choice is recurrent in topology optimization~\cite{filtrati,filtratiB}.
As far as all the parameters required by the optimization are concerned, we preserve the same values as in Sections~\ref{algo_sect}, while the computational mesh coincides with a $50\times50$ structured mesh.


\begin{figure}[hbt]
	\centering
	\includegraphics[width=1\linewidth]{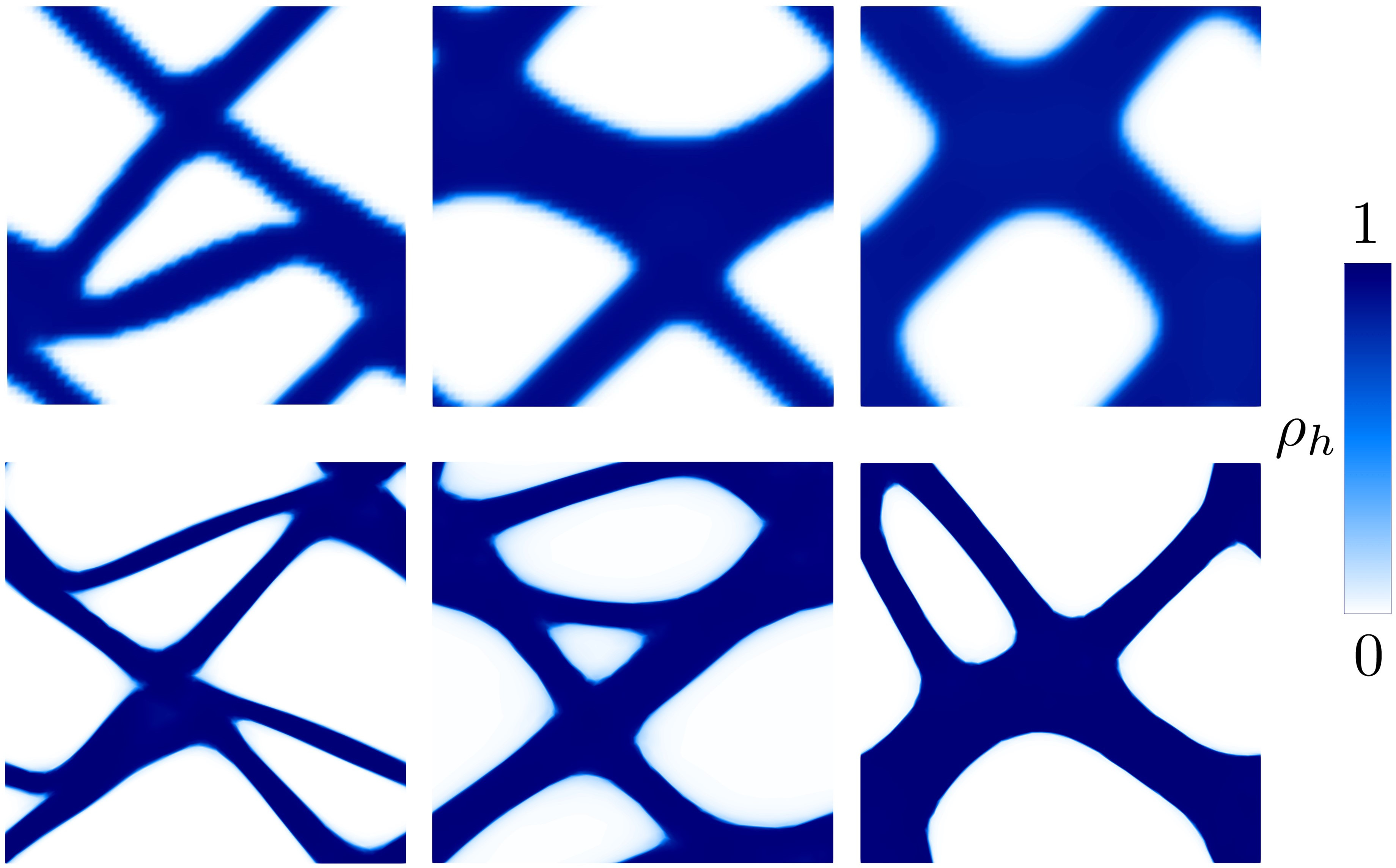}
	\caption{Comparison 
	between the optimized cells delivered by
	MultiP.microSIMP (top) and by
	MultiP-microSIMPATY   (bottom) for the Design cases $1$, $2$, $3$ (from left to right).}
	\label{approach_comparison}
\end{figure}


\begin{table}[bh]
	\begin{center}
		\caption{Comparison between 
		the optimized cells 
		delivered by MultiP-microSIMPATY and a standard inverse homogenization algorithm 
		in terms of mass.}
		\label{Tab1_comparison}
		\newcolumntype{D}{ >{\arraybackslash} m{3.75cm} }
		\newcolumntype{C}{ >{\centering\arraybackslash} m{0.8cm} }
		\begin{tabular}{l c c c}
			\hline
			& \cellcolor{Gray1Fill} D1 & \cellcolor{Gray2Fill} D2 & \cellcolor{Gray3Fill} D3 \\
			\hline
			MultiP-microSIMP & \cellcolor{Gray1Fill} 0.330 & \cellcolor{Gray2Fill} 0.443 & \cellcolor{Gray3Fill} 0.486 \\[1mm]
			MultiP-microSIMPATY & \cellcolor{Gray1Fill} 0.292 & \cellcolor{Gray2Fill} 0.412 & \cellcolor{Gray3Fill} 0.415 \\[1mm]
			\hline
			Mass reduction [\%] & \cellcolor{Gray1Fill} 11.5\% & \cellcolor{Gray2Fill} 7.0\% & \cellcolor{Gray3Fill} 14.6\% \\
			\hline
		\end{tabular}
	\end{center}
\end{table}

Figure~\ref{approach_comparison} compares the optimized layouts delivered by MultiP-microSIMP (top) and MultiP-microSIMPATY (bottom) for the three design cases in Section~\ref{results}.
The topologies characterizing the three cells vary when resorting to mesh adaptation. In general, MultiP-microSIMPATY provides more complex layouts, which however are still manufacturable. The presence of intermediate densities in the cells yielded by MultiP-microSIMP is highligthed by the blurred structure contours, promoted by the massive employment of filtering. Table \ref{Tab1_comparison} quantitatively assesses the optimization performance of the two algorithms, by collecting the mass of the corresponding unit cells, together with the percentage mass reduction ensured by MultiP-microSIMPATY. On average, a mass saving of approximately 10\% is guaranteed 
by the sharp detection of the material/void interface, i.e., by the removal of intermediate densities.

\begin{figure}[h!]
	\centering
    \includegraphics[width=0.8\linewidth]{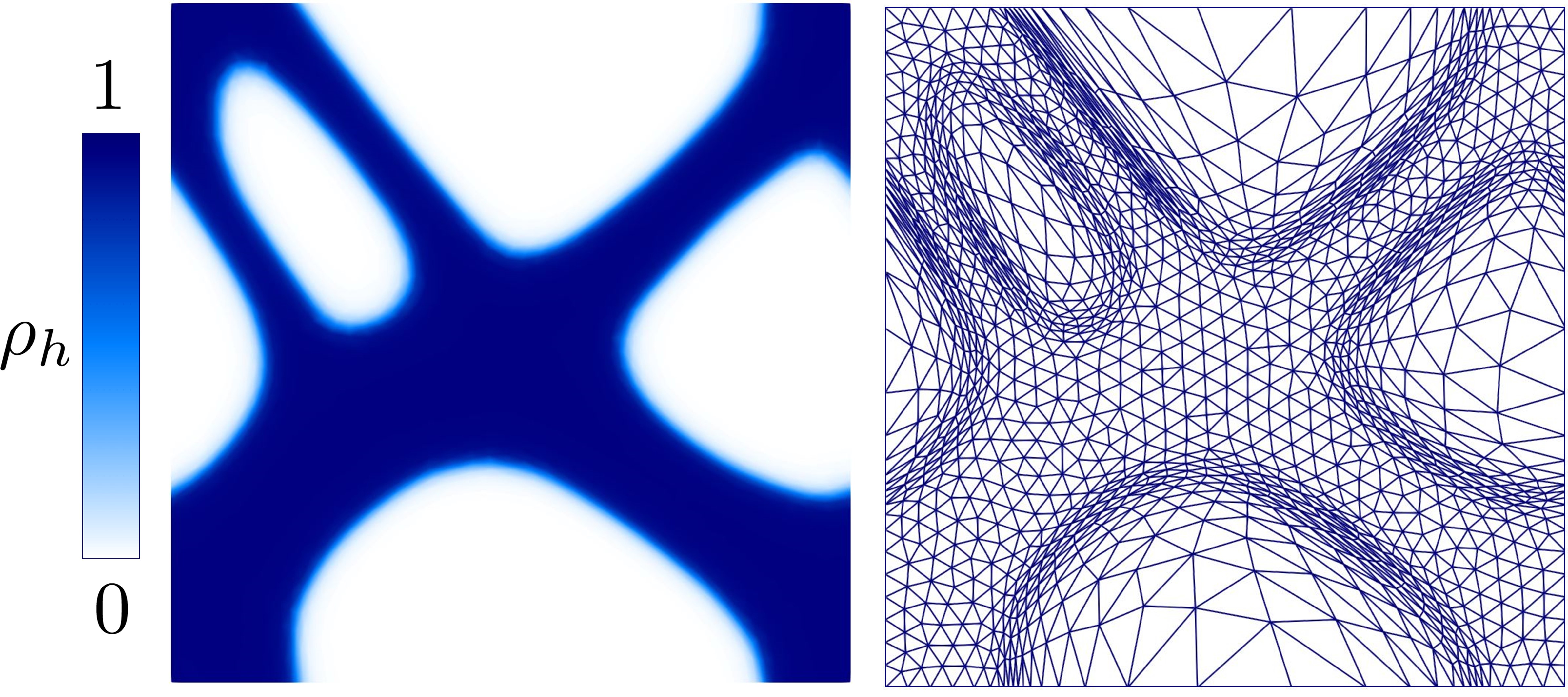}\\
    \includegraphics[width=0.8\linewidth]{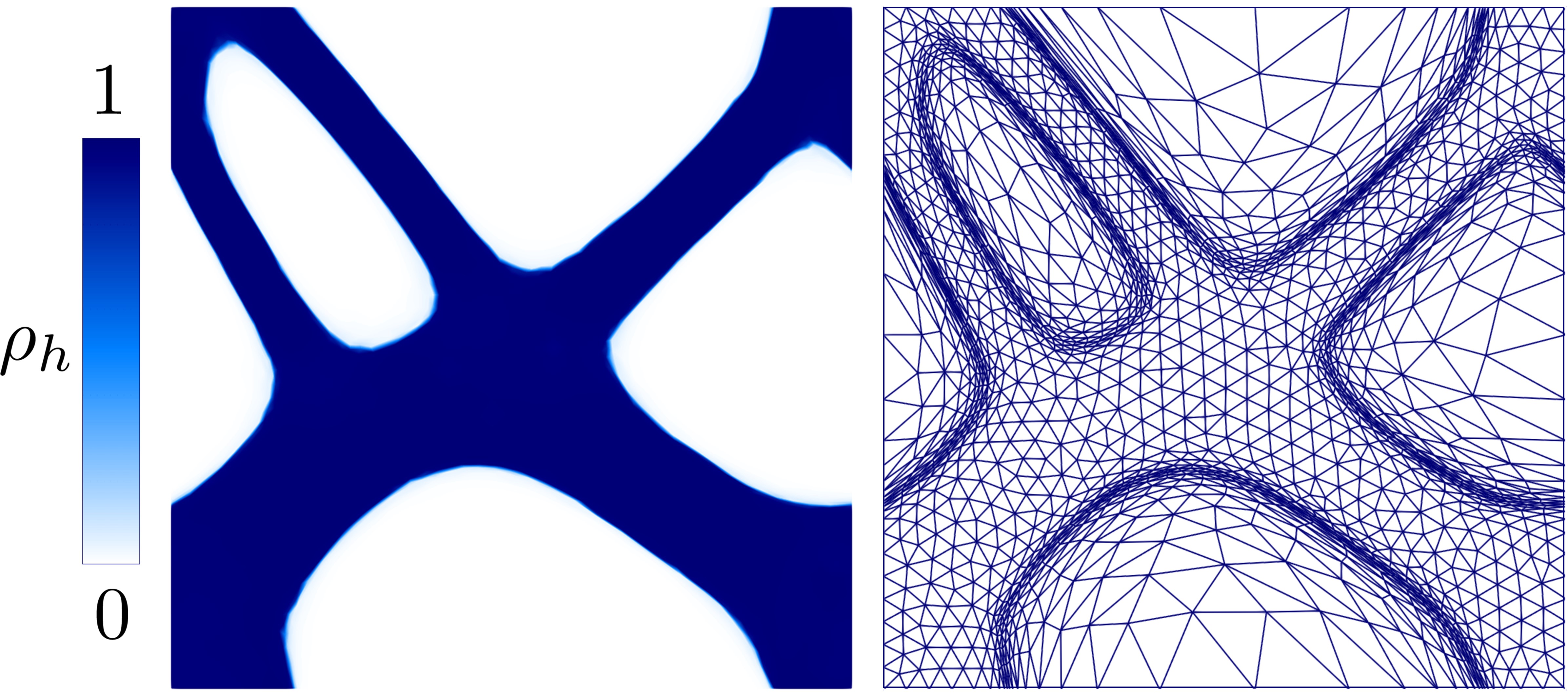}
	\caption{Effect of filtering for the MultiP-microSIMPATY algorithm: density field (left) and associated anisotropic adapted mesh (right) when filtering is applied during the whole  
	optimization process (top) and in the first $25$ iterations only (bottom).}
	\label{filter_effect}
\end{figure}

The use of filtering deserves further discussion. In particular, we prove the redundancy of the filtering phase after a sufficiently large number of global optimization iterations. To this aim, we run Algorithm~\ref{algo} for ${\tt kfmax} = 25$ and ${\tt kfmax} = {\tt kmax}$ (i.e., smoothing and sharpening filters in lines 6-7 are applied at each global iteration).
Figure~\ref{filter_effect} compares the output associated with these two choices. The final topology provided by both the procedures is the same. This confirms that filtering is instrumental only in the identification of the final layout, and this takes place during the first iterations. From the top-left panel, the slightly diffusive action of the selected filtering is also evident, giving rise to intermediate densities along the layout boundaries. On the other hand, the removal of filtering allows mesh adaptation to sharply detect gradients from material to void, thus increasing the quality of the final output (compare the two panels on the left panel). The improvement in terms of boundary detection is confirmed also by the final adapted mesh, which captures the steep gradients of the density with thinner refined areas (compare the two panels on the right).

Finally, we highlight that the presence of blurred interfaces may raise issues in the extraction of the final geometry, after the optimization procedure. In fact, the extracted geometry strongly depends on the cut-off threshold, with possible significant alteration of the overall mass and the expected thermo-mechanical properties.

\section{Conclusions and perspectives}\label{conclusions}
In this paper,
we provide a new methodology for the design of cellular materials optimized by means of multi-physics inverse homogenization, discretized on customized computational meshes.
The inverse homogenization problem is modeled by a standard density-based topology optimization at the microscale; the grid is generated by exploiting an anisotropic a posteriori error estimator which drives a mesh adaptation procedure. These two phases are iteratively coupled in the MultiP-microSIMPATY algorithm, in order to deliver layouts characterized by clear-cut contours.
In particular, goal of the analyzed test cases is the design of lightweight structures with prescribed elastic and thermal properties, according to a multi-physics framework.

The main results of this work can be outlined as follows:
\begin{itemize}
	\item[i)] MultiP-microSIMPATY algorithm provides original design solutions, complying also with conflicting requirements;
	\item[ii)] the good performance of microSIMPATY has been confirmed also in a multi-physics context. Standard issues typical of topology optimization, such as the presence of intermediate densities, of jagged boundaries, and of too complex structures is mitigated by the employment of a mesh customized to the design process (see Figure~\ref{approach_comparison} and Table~\ref{Tab1_comparison});
	\item[iii)] the new cellular materials have been successfully compared with consolidated solutions, in terms of mechanical and thermal properties (see Table~\ref{Tab2_comparison});
	\item[iv)] filtering can be considerably limited thanks to the use of mesh adaptation. This turns into an improvement in terms of accuracy of the optimization process (see Figure~\ref{filter_effect});
	\item[v)] the employment of an anisotropic mesh adaptation provides advantages with a view to a manufacturing phase. Indeed, the unit cells designed by MultiP-microSIMPATY exhibit very smooth geometries which demand for a very limited post-processing;
	\item[vi)] the procedure here settled turns out to be fully general with respect to the selected multi-physics context.
\end{itemize}

Possible future developments include the extension of the MultiP-microSIMPATY design procedure to a 3D setting. The proposed methodology could also be exploited in a multiscale topology optimization framework~\cite{Ferro2021}, inspired by the many possible applications in engineering practice (including medicine, aerospace, automotive, architecture).
In such a context, with a view to the manufacturing step, another issue which deserves further investigation is represented by the handling of the transition area between different cellular materials.
Finally, innovative techniques, such as model reduction or machine learning, still represent topics of high relevance in topology optimization for a future examination~\cite{oliver19,ferro19,mirabella21}

\section*{Acknowledgements}
This research is part of the activity of the METAMatLab at Politecnico di Milano. The first and the last authors acknowledge the Italian Ministry of Education, University and Research for the support provided through the "Department of Excellence LIS4.0 - Lightweight and Smart Structures for Industry 4.0” Project. The second author thanks Istituto Nazionale di Alta Matematica (INdAM) for the awarded grant.
Finally, the third author acknowledges the research project GNCS-INdAM 2020 “Tecniche Numeriche Avanzate per Applicazioni Industriali”.

\section*{Data availability}
Data are not available.

\biboptions{sort&compress}
\bibliographystyle{elsarticle-num}
\bibliography{main}

\end{document}